\documentclass[11pt]{article}
\usepackage{axodraw}
\usepackage{epsfig}
\usepackage{amsfonts}

\newcommand{\la}[1]{\label{#1}}
\newcommand{\be}{\begin{equation}}
\newcommand{\ee}{\end{equation}}
\newcommand{\ba}{\begin{eqnarray}}
\newcommand{\ea}{\end{eqnarray}}
\newcommand{\bi}{\begin{itemize}}
\newcommand{\ei}{\end{itemize}}
\newcommand{\rmi}[1]{{\mbox{\scriptsize #1}}}
\newcommand{\nr}[1]{(\ref{#1})}
\newcommand{\tr}{{\rm Tr\,}}

\newcommand{\nn}{\nonumber \\}
\newcommand{\fr}[2]{{\frac{#1}{#2}}}

\renewcommand{\vec}[1]{{\bf #1}}

\newcommand{\Nf}{N_{\rm f}}

\newcommand{\Nc}{N_{\rm c}}

\newcommand{\xpt}{$\chi$PT\ } 
\newcommand{\zz}%
 {{\mathbb{Z}}^4}
\renewcommand{\a}{r}    
\renewcommand{\b}{s}    
\renewcommand{\c}{u}    
\renewcommand{\d}{v}    
\newcommand{\cK}{{\cal K}}

\newcommand{\RR}{{\rm I\kern -.2em  R}}
\newcommand{\eq}{eq.~} 
\newcommand{\eqs}{eqs.~} 
\newcommand{\fig}{Fig.~}
\newcommand{\figs}{Figs.~}
\newcommand{\se}{Sec.~}

\def\lsi{\raise0.3ex\hbox{$<$\kern-0.75em\raise-1.1ex\hbox{$\sim$}}}
\def\gsi{\raise0.3ex\hbox{$>$\kern-0.75em\raise-1.1ex\hbox{$\sim$}}}

\newcommand{\gsim}{\mathop{\gsi}}
\newcommand{\hide}[1]{ }

\makeatletter \@addtoreset{equation}{section} \makeatother
\renewcommand{\theequation}{\arabic{section}.\arabic{equation}}
\makeatletter
\renewcommand\section{\@startsection {section}{1}{\z@}%
                                   {-5.5ex \@plus -1ex \@minus -.2ex}
                                   {2.3ex \@plus.2ex}%
                                   {\normalfont\large\bfseries}}
\renewcommand\subsection{\@startsection{subsection}{2}{\z@}%
                                     {-3.25ex\@plus -1ex \@minus -.2ex}%
                                     {1.5ex \@plus .2ex}%
                                     {\normalfont\normalsize\bfseries}}
\renewcommand\thesection {\@arabic\c@section}
\renewcommand\thesubsection   {\thesection.\@arabic\c@subsection}
\renewcommand{\@seccntformat}[1]{%
\csname the#1\endcsname.\hspace{1.0em}}
\makeatother

\input pix.sty

\begin{document}

\begin{titlepage}
\vspace{-1cm}
\begin{flushright}
\begin{tabular}{ll}
BI-TP 2008/04 & Edinburgh 2008/07 \\
FTUAM-08-01& FTUV-08-0226 \\
IFIC/08-07 &IFT-UAM-CSIC-08-07 \\
MKPH-T-08-01\\
\end{tabular}
\end{flushright}

\vspace{2cm}
\begin{centering}
\vfill


{\Large\bf Determination of the $\boldmath{\Delta S = 1}$ weak Hamiltonian in the SU(4)
\\[2mm] 
chiral limit through topological zero-mode wave functions}

\vspace*{0.8cm}

P.~Hern\'andez$^{\rm a}$, 
M.~Laine$^{\rm b}$, 
C.~Pena$^{\rm c}$,  
E.~Torr\'o$^{\rm a}$,  
J.~Wennekers$^{\rm d}$,  
H.~Wittig$^{\rm e}$
\vspace*{0.5cm}

{\em $^{\rm a}$%
Dpto.\ F\'{\i}sica Te\'orica, 
Universidad de Valencia and IFIC-CSIC \\ 
Apt.\ 22085, E-46071 Valencia, Spain\\}

\vspace{0.3cm}

{\em $^{\rm b}$%
Faculty of Physics, University of Bielefeld, 
D-33501 Bielefeld, Germany\\}

\vspace{0.3cm}

{\em $^{\rm c}$%
Dpto. F\'{\i}sica Te\'orica and Instituto de F\'{\i}sica Te\'orica UAM/CSIC\\ 
Facultad de Ciencias, Universidad Aut\'onoma de Madrid \\
Cantoblanco, E-28049  Madrid, Spain}

\vspace{0.3cm}

{\em $^{\rm d}$%
SUPA, School of Physics, University of Edinburgh, Edinburgh EH9 3JZ, UK}

\vspace{0.3cm}

{\em $^{\rm e}$%
Institut f\"ur Kernphysik, 
University of Mainz, 
D-55099 Mainz, Germany\\}

\vspace*{0.8cm}

{\bf Abstract}
 
\end{centering}
 
\vspace*{0.4cm}

\noindent
A new method to determine the low-energy couplings of the $\Delta S=1$ 
weak Hamiltonian is presented. It relies on a matching of the topological 
poles in $1/m^2$ of three-point correlators of two pseudoscalar densities and 
a four-fermion operator, measured in lattice QCD, to the same observables 
computed in the $\epsilon$-regime of chiral perturbation theory. 
We test this method in a theory with a light charm quark, 
i.e.\ with an SU(4) flavour symmetry. Quenched numerical 
measurements are performed in a 2$~$fm box, and chiral perturbation 
theory predictions are worked out up to next-to-leading order. 
The matching of the two sides allows to determine the weak 
low-energy couplings in the SU(4) limit. We compare the results with 
a previous determination, based on three-point correlators containing two 
left-handed currents, and discuss the merits and drawbacks 
of the two procedures. 
  


\vspace*{1cm}
 
\noindent

\vfill
 
\end{titlepage}


\section{Introduction}

Understanding quantitatively or even just qualitatively 
non-leptonic kaon decays, $K\to\pi\pi$, starting 
from the QCD Lagrangian, remains a formidable challenge despite
decades of work. 
The difficulty is that the non-perturbative low-energy dynamics 
of strong interactions plays an essential role~\cite{mk}, yet it has
turned out to be difficult to reduce the systematic errors of 
lattice studies of these effects~\cite{lat,b} to a tolerable level, 
because of the prohibitive cost of treating chiral symmetry, final 
state kinematics, and physical quark masses without compromises  
(for the current status and references, see ref.~\cite{Juttner:2007sn}).

A somewhat less challenging approach amounts to abandoning the direct computation of $K \to\pi\pi$ decay amplitudes in favour of determining, via lattice simulations, the low-energy couplings (LECs) of the effective chiral weak Hamiltonian that describes these decays \cite{b}. This can 
be achieved by matching lattice measurements of suitable correlation 
functions to the same correlation functions computed  within chiral 
perturbation theory ($\chi$PT). Among the simplifications thus achieved
are that the matching does not necessitate physical 
kinematics, and that physical quark masses are not needed either, 
as long as the regime of validity of $\chi$PT is reached. This however requires sufficiently large volumes and small quark masses. 

There are various possibilities
for the order in which the volume is increased and the quark masses are
decreased. In fact, it turns out to be useful to approach the chiral
limit by first decreasing the quark masses. The reason is that in this
parameter range, referred to as the $\epsilon$-regime of $\chi$PT~\cite{GL} 
(see also ref.~\cite{N}), it is possible to work out 
next-to-leading order corrections in $\chi$PT without introducing 
any more LECs than at the leading order, thereby putting the $\chi$PT
side of the matching well under control. 

At the same time, carrying out lattice simulations 
in the $\epsilon$-regime is quite demanding. Fortunately, 
the advent of Ginsparg-Wilson formulations of lattice 
fermions~\cite{gw}--\cite{kn}, which possess an exact chiral 
symmetry in the limit of vanishing quark masses, and many 
subsequent developments on the numerical side~\cite{methods}, 
have made such simulations possible. 

In ref.~\cite{strategy} a strategy based 
on these methods was proposed, with the goal of revealing the 
role that the charm quark mass plays in $K\to\pi\pi$ decays. 
The first step is the determination of the leading-order weak
LECs in a theory with a light charm quark, that is in
a four-flavour theory with an exact SU(4) chiral symmetry in the valence
sector. The first results of this computation, from simulations in the $\epsilon$-regime, have been presented
in ref.~\cite{prl}. The next step of the strategy is to increase the 
charm mass and monitor the LECs as we move towards a theory with 
an SU(3) flavour symmetry~\cite{strategy,largemc}.  

The observables used for performing the matching between lattice QCD 
and the chiral effective theory in ref.~\cite{prl} were three-point 
correlators of two left-handed currents and a weak operator.
(The $\chi$PT side for these observables has also been worked
out for SU(3)~\cite{weak}.)  
In this paper we pursue the same goal by means of a different
type of observable. Indeed, we propose to consider a correlation function 
of two pseudoscalar densities and a weak operator. The peculiarity of
this correlator is that it has poles in $1/m^2$ in 
the $\epsilon$-regime, when evaluated in sectors of 
non-vanishing topological charge (which we define {\`a la} ref.~\cite{hln}). As we will show, the residues of
these poles are easier to compute numerically than the correlation
functions themselves, since they typically require fewer quark
propagators, which are substituted by projectors on the zero-mode
wave functions. We then use the residues to perform the matching
between the fundamental and effective theories and 
determine the SU(4) weak LECs.

Let us stress that while the weak LECs themselves are universal, 
and can in principle be determined with any method, it is difficult 
to know {\em a priori} which of the multitude of possible 
strategies is the optimal one in practice.  
This depends, on one hand, on the numerical cost of the measurements 
involved, and on the other, on how well chiral perturbation theory
converges for the observable in question. We will compare the two 
methods mentioned (ref.~\cite{prl} and the present one)
on both accounts. The hope is that by carrying out this
comparison in the quenched approximation, it will become clear 
whether or not a particular strategy is superior  when one moves  
to the more expensive unquenched environment, or if it remains 
the best policy to probe the LECs by a combination
of independent techniques.  

The paper is organized as follows. 
In \se\ref{sec:theo} we introduce the observables to be computed 
in the fundamental theory and present the results of a next-to-leading 
order computation of the same observables in $\chi$PT, in the 
$\epsilon$-regime.  Besides the three-point functions previously mentioned, 
we will consider two-point functions that we use for normalization. 
In \se\ref{sec:num}, we present the results of a quenched numerical
computation of these amplitudes in a $2~$fm box and 
a new determination of the weak LECs in the SU(4) limit. 
We conclude in \se\ref{sec:conclu}.

\section{Low-energy couplings from zero-mode wave functions}
\label{sec:theo}

In the $\epsilon$-regime and in a fixed topological sector, correlation 
functions involving quark propagators may contain poles in $1/(mV)^n$, 
where $n$ is some integer, whenever the contribution of the zero-modes 
to the spectral representation of the quark propagator gives a non-vanishing 
contribution to the correlation function.  For a number of reasons the residues turn out to be easier to compute than the correlation functions themselves. The idea, 
explored in detail in ref.~\cite{zeromode}, is then to use the residues of 
the topological poles to perform the matching between QCD and $\chi$PT, 
instead of the full correlation function. Given a correlation function 
$C_\nu(x_1,x_2,...)$, the residue can be isolated by
\begin{eqnarray}
 C_{\nu}(x_1,x_2,...) \equiv {\mbox{Res}_n\over (m V)^n} + ...
 \;, \qquad
 \mbox{Res}_n = \lim_{m\rightarrow 0} (m V)^n C_{\nu}(x_1,x_2,...)
 \;. 
\end{eqnarray}

In ref.~\cite{zeromode} the two-point function of the pseudoscalar density 
was considered in this context. The presence of a pole in $1/(m V)^2$ implies 
that the corresponding residue can be computed fully in terms of the zero-mode
wave functions: no propagator computation is required. On the effective theory 
side, the same pole does appear and, up to a certain order, the residue is 
a function of only the pseudoscalar decay constant, $F$, and the volume
(in the quenched theory additional couplings appear).  An exploratory 
numerical study in the quenched approximation was presented and the principal 
usefulness of the method to extract the low-energy coupling $F$ was confirmed. 
Similar investigations have also been reported in refs.~\cite{zeromode2}.

In the present work, we extend this idea to the computation of three-point 
functions from which the weak low-energy constants can be determined. 
In particular, we consider three-point functions of two pseudoscalar 
densities and a weak four-fermion operator. It is easy to see that 
such correlation functions do have poles in $1/(mV)^2$ when computed 
in non-trivial topological sectors in the $\epsilon$-regime, 
as we now show. 

\subsection{Correlators in the fundamental theory}

Following the strategy of refs.~\cite{strategy, prl}, 
we consider a theory with four light and degenerate flavours such 
that $m_u=m_d=m_s=m_c=m$, which we refer to as the GIM limit. 
After integrating out the $W^\pm$ to first order in the weak coupling, $g^2_{\rm{w}}$,  
the resulting Weak Hamiltonian is given by
\begin{eqnarray}
  {H}_{\rm w}=\frac{g^2_{\rm{w}}}{4M_W^2}V_{us}^*V_{ud}
  \left\{ 
       k_1^{+}~Z_{11}^+{Q}_1^{+} + k_1^{-}~Z_{11}^-~{Q}_1^{-}\right\}
  \;, 
  \label{eq_Hw_final}
\end{eqnarray}
where the operators ${Q}_1^\pm$ transform in the $\mathbf{84}$ 
and $\mathbf{20}$ representations of SU(4):
\begin{eqnarray}
& & {Q}_1^{\pm} = ([{O}_1]_{suud}\pm[{O}_1]_{sudu})
                      -(u\to c)
   \;,
    \label{eq_Q1pm} \\
& & [{O}_1]_{\a\b\c\d}\equiv
    \big(\bar{\Psi}_\a\gamma_\mu P_{-}\tilde{\Psi}_\c\big)
    \big(\bar{\Psi}_\b\gamma_\mu P_{-}\tilde{\Psi}_\d\big)
    \;.
    \label{eq_O1}
\end{eqnarray}
Here $k_1^\pm$ are Wilson coefficients at the scale $M_W$,  
and $Z^\pm_{11}$ are the corresponding renormalization factors.  
We will follow the renormalization prescription of refs.~\cite{strategy, prl},
that is we will use the RGI scheme, in which these factors are known 
non-perturbatively~\cite{renorm}. For any unexplained details we refer
the reader to these references.

Deep in the non-perturbative regime this effective Hamiltonian admits an expansion in terms of the Goldstone boson fields and can be represented as 
\begin{eqnarray}
  \mathcal{H}_{\rm w}=\frac{g^2_{\rm{w}}}{4M_W^2}V_{us}^*V_{ud}
  \left\{ 
       g_1^{+}~\mathcal{Q}_1^{+} + g_1^{-}~\mathcal{Q}_1^{-} + ...\right\}
 \;, 
\label{eq_Hw_xpt}
\end{eqnarray}
where $\mathcal{Q}_1^\pm$ are operators made out of the Goldstone field, 
and terms of higher order in the chiral expansion
have been omitted. Our task is to match for the coefficients $g_1^\pm$
in the chiral limit, by comparing lattice simulations with $\chi$PT
predictions. 

Now, given 
that QCD dynamics itself respects chiral symmetry, the results of such a
matching are independent of the precise flavour indices appearing in 
\eq\nr{eq_Q1pm}, as long as the operators remain traceless and have
the correct symmetry properties. In practice, it is indeed convenient to 
consider the operators
\be
 O_1^\pm \equiv [{O}_1]_{\a\b\c\d}\pm[{O}_1]_{\a\b\d\c}
 \;, \la{O1_def}
\ee
with {\em all indices different}, rather than 
$Q_1^\pm$; these operators are automatically
traceless, and no subtraction of the type in \eq\nr{eq_Q1pm} is needed.
The results for the matching of $g_1^\pm$
are nevertheless guaranteed to be identical.

We want to carry out the matching 
in a finite volume through the computation of the following 
bare three-point functions in a fixed topological sector of charge $\nu$:
\begin{eqnarray}
 A_\nu^\pm (x_0-z_0, y_0-z_0) \equiv  - \lim_{m\rightarrow 0} 
 (mV)^2 \int_{\vec{x}} \int_{\vec{y}}\langle \partial_{x_0} P^a(x) 
 O_1^\pm(z) \partial_{y_0} P^b(y)\rangle_\nu
 \;, \la{3pt_qcd}
\end{eqnarray}
where the bare pseudoscalar density
reads $P^a \equiv i \bar{\Psi} \gamma_5 T^a \tilde\Psi$, and $m$
is the bare quark mass.  
These amplitudes get contributions from two possible contractions, 
colour-disconnected, ${\bar A}_\nu$, and colour-connected, ${\tilde A}_\nu$:
\begin{eqnarray}
 A_\nu^\pm 
 =
 (T^a_{ur}T^b_{vs}+T^a_{vs}T^b_{ur}\pm T^a_{vr}T^b_{us}\pm T^a_{us}T^b_{vr}) 
 \Bigl[ {\bar A}_\nu 
 \pm 
 {\tilde A}_\nu  
 \Bigr]
 \;, \la{Anu_pm}
\end{eqnarray}
where
\ba
 {\bar A}_\nu 
 \!\! & \equiv & \!\!
  \lim_{m\rightarrow 0} 
 (mV)^2 \int_{\vec{x}} \int_{\vec{y}}\partial_{x_0} \partial_{y_0}
  \tr[S_m(x,z) \gamma_\mu P_- S_m(z,x) \gamma_5]\,
  \tr[S_m(y,z) \gamma_\mu P_- S_m(z,y) \gamma_5]
 \;, \nn  
 {\tilde A}_\nu 
 \!\! & \equiv & \!\!
  - \lim_{m\rightarrow 0} 
 (mV)^2 \int_{\vec{x}} \int_{\vec{y}}\partial_{x_0} \partial_{y_0}
  \tr[S_m(x,z) \gamma_\mu P_- S_m(z,y) \gamma_5
  S_m(y,z) \gamma_\mu P_- S_m(z,x) \gamma_5]
 \;. \hspace*{5mm} \nn \la{Anu_expl}
\ea
Here $S_m$ is the massive quark propagator. 

It is convenient to normalize these three-point functions 
with bare two-point functions of the form
\ba
 -i \tr[T^a T^b] B_\nu(x_0-z_0) & \equiv &  
 \lim_{m\rightarrow 0} (mV) \int_{\vec{x}} \langle \partial_{x_0} P^a(x)
 {L}_0^b(z) \rangle_\nu
 \;, \la{B_nu}
\ea
where the bare left current reads 
${L}_0^a \equiv \bar{\Psi} \gamma_0 P_- T^a \tilde{\Psi}$.  
Carrying out the contractions, we get
\ba
 B_\nu(x_0-z_0)
  & = & 
 \lim_{m\rightarrow 0} (mV) \int_{\vec{x}} \partial_{x_0}
 \tr[S_m(x,z) \gamma_0 P_- S_m(z,x) \gamma_5]
 \;. \la{Bnu_expl}
\ea
Note that the two-point function in \eq\nr{B_nu} can be related 
through the non-singlet axial Ward identity to the 
two-point function of two pseudoscalar densities, 
considered in ref.~\cite{zeromode}:
\ba
 \tr[T^a T^b] Z_A B_\nu(x_0-z_0) & = &
 i \lim_{m\rightarrow 0} (mV) \partial_{x_0} \int_{\vec{x}} \langle  P^a(x)
 Z_A {L}_0^b(z) \rangle_\nu 
 \nn & = & 
 -i \lim_{m\rightarrow 0} (mV) \partial_{z_0} \int_{\vec{z}} \langle  P^a(x)
 {Z_A L}_0^b(z) \rangle_\nu 
 \nn & = & 
 \lim_{m\rightarrow 0} (m^2 V)  
 \int_{\vec{x}} \langle P^a(x)~ P^b(z)\rangle_\nu
 \;, 
 \label{wi}
\ea
where  $Z_A$ denotes  the renormalization constant of the currents $L_0^a$.
Here we made use of the fact that the product $m P^a$ does not 
require renormalization. 

Now, let us see why it is useful to consider the quantities
in \eqs\nr{Anu_expl}, \nr{Bnu_expl}. 
The point is that $\bar A_\nu, \tilde A_\nu$ and $B_\nu$ 
are zero, unless some of the quark propagators are saturated by zero modes. 
Let us denote by $v_i(x) \in \cK$ the zero-mode wave functions 
(recall that the dimension of the kernel of the Dirac operator 
is  $\mbox{dim}(\cK)=|\nu|$) that are normalized as
\begin{equation}
 \int_{x} \, v_i^\dagger(x) v_i(x) = V 
 \;.
\end{equation}
The spectral representation of the quark propagator then reads
\be
 S_m(x,y) = \sum_{i=1}^{|\nu|}
 \frac{v_i(x) v_i^\dagger (y)}{mV} + ...
 \;. 
\ee
We can define the sources
\begin{eqnarray}
 {\eta}_i(z;x_0) & \equiv &  
 \partial_{x_0} \int_{\vec{x}}  P_{-\chi} S_m(z,x) P_\chi v_i(x)
 \;, \nn 
 {\eta}^\dagger_i(z;x_0) & = &  
 - \partial_{x_0} \int_{\vec{x}}  v_i^\dagger(x) P_{\chi} S_m(x,z) P_{-\chi} 
 \;,  
\end{eqnarray}
where $\chi$ is the chirality of the zero-modes. Given that
$\gamma_\mu P_- = P_+ \gamma_\mu P_-$, we note that, depending 
on chirality, only some of the propagators can be saturated with 
zero-modes: for $\nu < 0$, the ones multiplying $P_-$, and for
$\nu > 0$, the ones multiplying $P_+$. If $\nu > 0$, 
the disconnected and connected amplitudes of the three-point functions
thus become
\begin{eqnarray}
 \bar{A}_\nu(x_0-z_0,y_0-z_0) & = & 
 \lim_{m\rightarrow 0}~{1 \over L^3}~\int_{\vec{z}}~ \Bigl\langle 
 \sum_{i=1}^{|\nu|} v_i^\dagger(z) \gamma_\mu \eta_i(z;x_0) 
 \sum_{j=1}^{|\nu|} v_j^\dagger(z) \gamma_\mu \eta_j(z;y_0) 
 \Bigr\rangle_\nu 
 \;,
 \nonumber  \\ 
 \tilde{A}_\nu(x_0-z_0,y_0-z_0) & = & 
 - \lim_{m\rightarrow 0}~{1 \over L^3}~\int_{\vec{z}} \Bigl\langle 
 \sum_{i,j =1}^{|\nu|} v_i^\dagger(z) \gamma_\mu \eta_j(z;y_0) 
 v_j^\dagger(z) \gamma_\mu \eta_i(z;x_0) 
 \Bigr\rangle_\nu 
 \;, 
 \label{aatildep}
\end{eqnarray}
while for $\nu <0$ we get
\begin{eqnarray}
 \bar{A}_\nu(x_0-z_0,y_0-z_0) & = & 
 \lim_{m\rightarrow 0}~{1 \over L^3}~\int_{\vec{z}}~ \Bigl\langle 
 \sum_{i =1}^{|\nu|} \eta_i^\dagger(z;x_0) \gamma_\mu v_i(z) 
 \sum_{j =1}^{|\nu|} \eta_j^\dagger(z;y_0) \gamma_\mu v_j(z) 
 \Bigr\rangle_\nu 
 \;,
\nonumber  \\ 
 \tilde{A}_\nu(x_0-z_0,y_0-z_0) & = & 
 - \lim_{m\rightarrow 0}~{1 \over L^3}~\int_{\vec{z}} 
 \Bigl\langle 
 \sum_{i,j=1}^{|\nu|} \eta_i^\dagger(z;x_0) \gamma_\mu v_j(z) 
 \eta_j^\dagger(z;y_0) \gamma_\mu v_i(z) 
 \Bigr\rangle_\nu  
 \;. \label{aatildem}
\end{eqnarray}
For the two-point function of \eq\nr{Bnu_expl}, 
the positive chirality case 
$\nu > 0$ yields
\begin{eqnarray}
 B_\nu(x_0-z_0) = 
 \lim_{m\rightarrow 0}~\frac{1}{L^3} \int_{\vec{z}}
 \Bigl\langle \sum_{i=1}^{|\nu|}
 v_i^\dagger(z) \gamma_0 {\eta}_i(z;x_0) 
 \Bigr\rangle_\nu 
 \;, \label{bp}
\end{eqnarray}
while for $\nu < 0$ we arrive at
\begin{eqnarray}
 B_\nu(x_0-z_0) = 
 \lim_{m\rightarrow 0}~
 \frac{1}{L^3} \int_{\vec{z}}
 \Bigl\langle \sum_{i=1}^{|\nu|}
 \eta_i^\dagger(z;x_0) \gamma_0 v_i(z) 
 \Bigr\rangle_\nu 
 \;.
 \label{bm}
\end{eqnarray}
Finally, the Ward identity of eq.~(\ref{wi}) implies
\begin{eqnarray}
 Z_A B_\nu(x_0-z_0) = D_\nu(x_0-z_0) \equiv 
 \frac{1}{V} \sum_{i,j = 1}^{|\nu|} \int_{\vec{x}} 
 \left\langle v_j^\dagger(x) v_i(x) v_i^\dagger(z) v_j(z) \right\rangle_\nu
 \;. \la{wi_expl}
\end{eqnarray}
Here the limit $m\rightarrow 0$ has been taken analytically on the right-hand
side, while it needs to be taken numerically on the left-hand side
(cf.\ \eqs\nr{bp}, \nr{bm}); therefore
\eq\nr{wi_expl} provides  a non-trivial test on our ability to approach 
the limit needed in \eqs\nr{aatildep}--\nr{bm}. 

It is clear from eqs.~(\ref{aatildep})--(\ref{bm}) that a number 
of inversions equal to twice the topological charge, i.e. $2 |\nu|$ 
(since $x_0$ and $y_0$ need to  be fixed), is sufficient for constructing 
the correlation functions, whilst averaging over all the spatial positions 
of the three sources. When employing the method of refs.~\cite{strategy,prl}, which is based on the left-handed current, the summation over the spatial positions of the three sources is only possible through low-mode averaging (LMA) \cite{current,dgs}, and only for the contributions of the low modes.
 The price of LMA is 
$12+2\times N_\rmi{low}$ inversions, where $N_\rmi{low}$ was 
the number of low modes treated separately. Typically $N_\rmi{low}$ can 
be as large as 20 for 2~fm boxes, and hence the numerical cost 
can be quite substantial.

On the other hand, if the low modes of the Dirac operator are 
known, as for example would be the case if low-mode preconditioning \cite{methods} is used, it is possible to  perform an additional averaging over time translations 
for the low-mode contributions to the correlation functions defined above. It is important to stress however that this extra low-mode averaging does not involve any additional inversion, therefore {\it the overhead is not proportional to $N_\rmi{low}$}, as in the standard case \cite{current}.
We will describe briefly how this works in 
section \ref{se:lma}.
  
Summarizing, in order to perform the matching between the fundamental weak 
Hamiltonian and the effective one,  we will be considering 
the bare ratios
\begin{eqnarray}
 {R}_\nu^\pm &\equiv & 
 \!\!\! 
 { \bar{A}_\nu(x_0-z_0,y_0-z_0) \pm {\tilde A}_\nu(x_0-z_0,y_0-z_0) 
 \over B_\nu(x_0-z_0) B_\nu(y_0-z_0) } 
 \;. \label{rnu}
\end{eqnarray}
The renormalized ratios needed for matching the  
LECs in \eq\nr{eq_Hw_xpt} are then obtained by 
multiplying these correlators by the renormalization 
factors ${Z_{11}^\pm / Z_A^2}$ (cf.\ \eq\nr{match} below);
the procedure is identical to the one in ref.~\cite{prl}
and we refer to that reference for details.

\subsection{Correlators in chiral perturbation theory}

We now present the results for the observables just defined 
in the Chiral Effective Theory.

\subsubsection{Two-point function}

The two-point correlation function that needs to be computed 
in $\chi$PT, corresponding to \eq\nr{B_nu},  is given by
\begin{equation}
 -i \tr[T^a T^b] \mathcal{B}_\nu(x_0-z_0)
 \equiv
 \lim_{m\rightarrow 0} (m V)  \partial_{x_0} \int_{\vec{x}} 
 \langle \mathcal{P}^a(x)~\mathcal{J}_0^b(z)\rangle_\nu  
 \;,
 \la{2pt_xpt}
\end{equation}
where 
\begin{eqnarray}
 \mathcal{J}_\mu^a &=& 
 \frac{F^2}{2} \tr\left[ T^a U \partial_\mu U^\dagger \right] + ... 
 \;, \label{eq:j} \\
 \mathcal{P}^a &=& i \frac{\Sigma}{2} \tr \left[ T^a (U - U^\dagger)\right] 
 + ... 
 \;, \label{eq:p}
\end{eqnarray}
and contact terms have been omitted from \eq\nr{2pt_xpt}. 

Since the mass is taken to zero in \eq\nr{2pt_xpt}, 
the computation is carried out according to the rules of 
the $\epsilon$-expansion~\cite{GL}. We work up to  next-to-leading
order (NLO). The results for the contributions of the individual graphs, 
as well as the various zero-mode and spacetime integrals appearing, 
are listed in appendix~A.

\begin{figure}[!t]
\begin{center}


 \epsfysize=7.5cm\epsfbox{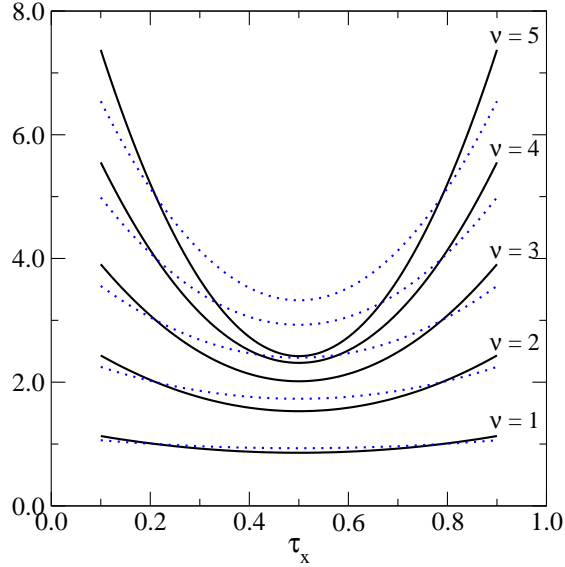}%

\caption{The two-point correlation function 
$T \mathcal{B}_\nu(x_0 - z_0)$ for $\Nf=2$ (solid) and $\Nf=0$ 
(dotted) as a function of $\tau_x = (x_0-z_0)/T$, for $T=L=2$ fm. 
The pion decay constant $F$ has been fixed to 93~MeV 
for $\Nf=2$ and to 110~MeV for $\Nf=0$.}
\label{fig:jp}
\end{center}
\end{figure}

Defining $\tau_x \equiv (x_0 - z_0)/T$ and $\rho \equiv T/L$,
and considering the unquenched theory,  
the final result from \eq\nr{B_complete} becomes, after replacing
$E(x)=G(x)/\Nf$ and using \eq\nr{st_A_2}, 
\begin{eqnarray}
 T \mathcal{B}_\nu(x_0-z_0) & = &  |\nu| 
 \biggl\{ 1 + \frac{2 \rho}{(FL)^2} \Bigl(|\nu| + \frac{1}{\Nf} \Bigr)
 h_1(\tau_x)
 \biggr\}
 \;,  \la{Bnu_res}
\end{eqnarray}
where 
\be
 h_1(\tau) \equiv  \frac{1}{2}  
 \left\{ \left[(\tau \mathop{\mbox{mod}} 1) 
 - {1 \over 2}\right]^2 - {1 \over 12}\right\}
 \;. \la{h1}
\ee
In \fig\ref{fig:jp} we show this result
for different values of $|\nu|$ in a symmetric box of size $T=L=2$~fm. 
The LO results are constants at $|\nu|$, so that all time dependence
results from the subleading corrections. 

In the quenched case we consider the $\epsilon$-counting 
described in ref.~\cite{zeromode}, introduced in order to set up 
a parametrically convergent perturbative series. As explained there, 
three new couplings can in principle appear at NLO: $\alpha, m_0^2, K$. 
The effect of $\alpha$ is to replace $1/\Nf \to \alpha/2\Nc$; 
however, this contribution is suppressed in 
the counting of ref.~\cite{zeromode} and will be omitted. 
Similarly, contributions from $m_0^2$ are also
suppressed and omitted. On the other hand, effects from the 
coupling $K$ could be of order unity, and need to be studied explicitly.  

The coefficient $K$ has two main effects. First of all, 
the pseudoscalar density is modified to
 \begin{eqnarray}
 P^a &=&  i {\Sigma \over 2} {\rm Tr}\left[ T^a (U - U^\dagger)\right] 
 - K \Phi_0 {\rm Tr}\left[ T^a (U+ U^\dagger)  \right] + ... 
 \;, \la{Pa_q}
\end{eqnarray}
where we have for brevity kept the notation of the unquenched theory, 
usable in the replica formulation of quenched $\chi$PT~\cite{ds}; 
for the notation in the supersymmetric formulation~\cite{BGS}, 
see \eq(3.2) of ref.~\cite{zeromode}.
In \eq\nr{Pa_q},  $\Phi_0 = {F \over 2} {\rm Tr}\left[ -i \ln U \right]$ 
is the singlet field. Second, the zero-mode integration measure 
is modified:
\begin{eqnarray}
 \langle ... \rangle^q_\nu = { \int_{ \rm U_0 \in U(N)}
 (...)~{\det}^\nu U_0 \, \,\exp\left[
 {\, \frac{\mu}{2} \tr (U_0+U_0^\dagger) + 2 \nu \frac{m K \Nc}{m_0^2 F}
 \tr(U_0 - U_0^\dagger)
 }\right]   \over \int_{\rm U_0 \in U(N)} 
 {\det}^\nu U_0 \,
 \exp\left[
 {\, \frac{\mu}{2} \tr (U_0+U_0^\dagger) + 2 \nu \frac{m K \Nc}{m_0^2 F}
 \tr(U_0 - U_0^\dagger)
 }\right]  }
 \label{Znu} 
 \;.
\end{eqnarray}
To first order in $K$, the effects come from 
the standard tree-level contribution computed with the 
modified weight of \eq\nr{Znu}, and from a new tree-level
term containing the $K$-term from \eq\nr{Pa_q}. 

The results for the two new contributions are given in 
\eqs\nr{HTopo_q_first}, \nr{HTopo_q_last} of appendix~A.
However, inserting the zero-mode integrals 
from \eqs\nr{HTopo_q_zero_first}, \nr{HTopo_q_zero_last}, 
these contributions cancel against each other. 
Therefore, 
\begin{eqnarray}
 T \mathcal{B}_\nu^q(x_0-z_0) & = &  |\nu| 
 \biggl\{ 1 + \frac{2 \rho |\nu| }{(FL)^2} h_1(\tau_x)  
 \biggr\}
 \;,  \la{Bnu_q_res}
\end{eqnarray}
and the result is very close to that in the full theory
(were it not that $F$ is different). Some examples 
are shown in \fig\ref{fig:jp}.

\subsubsection{Three-point function}

Next we consider the three-point function corresponding
to \eq\nr{3pt_qcd} in \xpt: 
\begin{equation}
 \mathcal{A}_\nu^\pm (x_0-z_0, y_0-z_0) \equiv  - \lim_{m\rightarrow 0} 
 (mV)^2 \int_{\vec{x}} \int_{\vec{y}}\langle 
 \partial_{x_0} \mathcal{P}^a(x) 
 \mathcal{O}_1^\pm(z) 
 \partial_{y_0} \mathcal{P}^b(y)
 \rangle_\nu
 \,. 
 \la{3pt_xpt}
\end{equation}
Here, like in \eq\nr{O1_def}, 
\be
 \mathcal{O}_1^\pm \equiv [\mathcal{O}_1]_{\a\b\c\d}
 \pm[\mathcal{O}_1]_{\a\b\d\c}
 \;, \la{O1_def_xpt}
\ee
where the weak operators are given by 
\begin{eqnarray}
 \mathcal{O}_{rsuv} = 
 {F^4 \over 4}~\left( \partial_\mu U  U^\dagger \right)_{ur}  
 \left( \partial_\mu U  U^\dagger \right)_{vs} + ...
 \;.
 \la{Orsuv_def_xpt}
\end{eqnarray}
Again, contact terms have been omitted from \eq\nr{3pt_xpt}.

Like for the two-point function, we work up to NLO. 
The results for the contributions of the individual graphs, 
as well as the various zero-mode and spacetime integrals appearing, 
are listed in appendix~B.

The result obtained after summing together all the graphs can be 
written as 
\begin{eqnarray}
 \mathcal{A}_\nu^\pm 
 =
 (T^a_{ur}T^b_{vs}+T^a_{vs}T^b_{ur}\pm T^a_{vr}T^b_{us}\pm T^a_{us}T^b_{vr}) 
 \Bigl[ {\bar \mathcal{A}}_\nu 
 \pm 
 {\tilde \mathcal{A}}_\nu  
 \Bigr]
 \;, \la{Anu_pm_xpt}
\end{eqnarray}
where, inserting $E(x)=G(x)/\Nf$ into \eq\nr{A_complete_2} and 
using \eq\nr{eq:GGGint}, one obtains 
\ba
 & & \hspace*{-1.5cm}
 {\bar \mathcal{A}}_\nu 
 \pm 
 {\tilde \mathcal{A}}_\nu  =  
  \Bigl( 1 \mp \frac{1}{|\nu|} \Bigr)
 \biggl\{
   \mathcal{B}_\nu(x_0-z_0)  
   \mathcal{B}_\nu(y_0-z_0)
 \nn & \pm &  
 \frac{\nu^2}{F^2 V}
 \Bigl[ 
  2 \beta_1 \rho^{-\fr32} 
 + 
  f_1(\tau_x) + f_1 (\tau_y) - h_1(\tau_x) - h_1(\tau_y) +
 \Bigl(1 \mp \frac{2}{\Nf} \Bigr) H(\tau_x,\tau_y) 
 \Bigr]  
 \biggr\}
 \;.  \hspace*{0.5cm} \la{A_full}
\ea
%
%
Here $\tau_x \equiv (x_0-z_0)/T$, $\tau_y \equiv (y_0-z_0)/T$, 
$\rho\equiv T/L$, 
$\beta_1$ is a \emph{shape coefficient} depending on the value of 
$\rho$~\cite{hal,h}, the function $h_1$ is defined in \eq\nr{h1}, and 
\ba 
 H(\tau_x,\tau_y) & \equiv & 
 h_1'(\tau_x) h_1'(\tau_y)   - h_1(\tau_x-\tau_y)  
 - \Bigl[ h_1'(\tau_x) -h_1'(\tau_y) \Bigr] h_1'(\tau_x-\tau_y)
 \;, \la{Hxy} \\ 
 f_1(\tau) & \equiv &   \left[h'_1(\tau)\right]^2 + 
 \sum_{\vec{p} \neq 0} \left[{ |\vec{p}|^2 C_\vec{p}(\tau)^2 
 + C'_\vec{p}(\tau)^2 }\right] 
 \;. \la{f1}
\ea
Furthermore, $\vec{p} = 2 \pi \rho\, \vec{n}$ 
with $\vec{n} = (n_1, n_2, n_3)$ being a vector of integers, and 
\begin{eqnarray}
 C_\vec{p}(\tau) \equiv 
 {\cosh\left\{ |\vec{p}| \left[(\tau  \mathop{\mbox{mod}} 1 )
 -{1 \over 2}\right]\right\} 
  \over 2 |\vec{p}| \sinh(|\vec{p}|/2)}
 \;, \la{Cp} \quad
 C_\vec{p}'(\tau) = 
 {\sinh\left\{ |\vec{p}| \left[(\tau  \mathop{\mbox{mod}} 1)
 -{1 \over 2}\right]\right\} 
  \over 2  \sinh(|\vec{p}|/2)}
  \;. 
\end{eqnarray}

%
%

The first term in \eq\nr{A_full} has the form of a factorized 
contribution. We can cancel out this term by taking the ratio of 
the three-point function with respect to the product of two 
two-point functions, in analogy with \eq\nr{rnu}: 
\ba
 \mathcal{R}^\pm_\nu & \equiv &
 \frac{ {\bar \mathcal{A}}_\nu (x_0-z_0, y_0-z_0) 
  \pm 
        {\tilde \mathcal{A}}_\nu(x_0-z_0, y_0-z_0)}
  {\mathcal{B}_\nu(x_0-z_0)  
   \mathcal{B}_\nu(y_0-z_0)}
 \la{Rpm_xpt_def} \\ 
 & \equiv & 
 \Bigl( 1 \mp \frac{1}{|\nu|} \Bigr)
 \Bigl[ 1 \pm r_\pm(z_0) \Bigr]
 \;.  \la{R_r}
\ea
We have adopted a notation here where $x_0$, $y_0$ are assumed fixed, 
so that $r_\pm$ is a function of $z_0$ only. Inserting the tree-level
result $\mathcal{B}_\nu = |\nu|/T$ in the numerators when dividing 
the NLO correction, we then obtain from \eq\nr{A_full} that
\be
 r_\pm(z_0) = 
 \frac{\rho}{(FL)^2}
 \Bigl[ 
  2 \beta_1 \rho^{-\fr32} 
 + 
  f_1(\tau_x) + f_1 (\tau_y) - h_1(\tau_x) - h_1(\tau_y) +
 \Bigl(1 \mp \frac{2}{\Nf} \Bigr) H(\tau_x,\tau_y) 
 \Bigr]  
 \;. \la{rpm_full}
\ee
Note that $r_\pm$ is independent of $\nu$, i.e.\  the topology and 
volume dependences have completely factorized at this order. 
%
%
The low-energy couplings $g_1^{\pm}$ can now be obtained from the matching
\be
 g_1^\pm \mathcal{R}_\nu^\pm = k_1^\pm~ {Z_{11}^\pm \over Z_A^2} ~R_\nu^\pm
 \;, \la{match}
\ee
at sufficiently large distances between the sources. 

\begin{figure}[!t]
\begin{center}

 \epsfysize=7.5cm\epsfbox{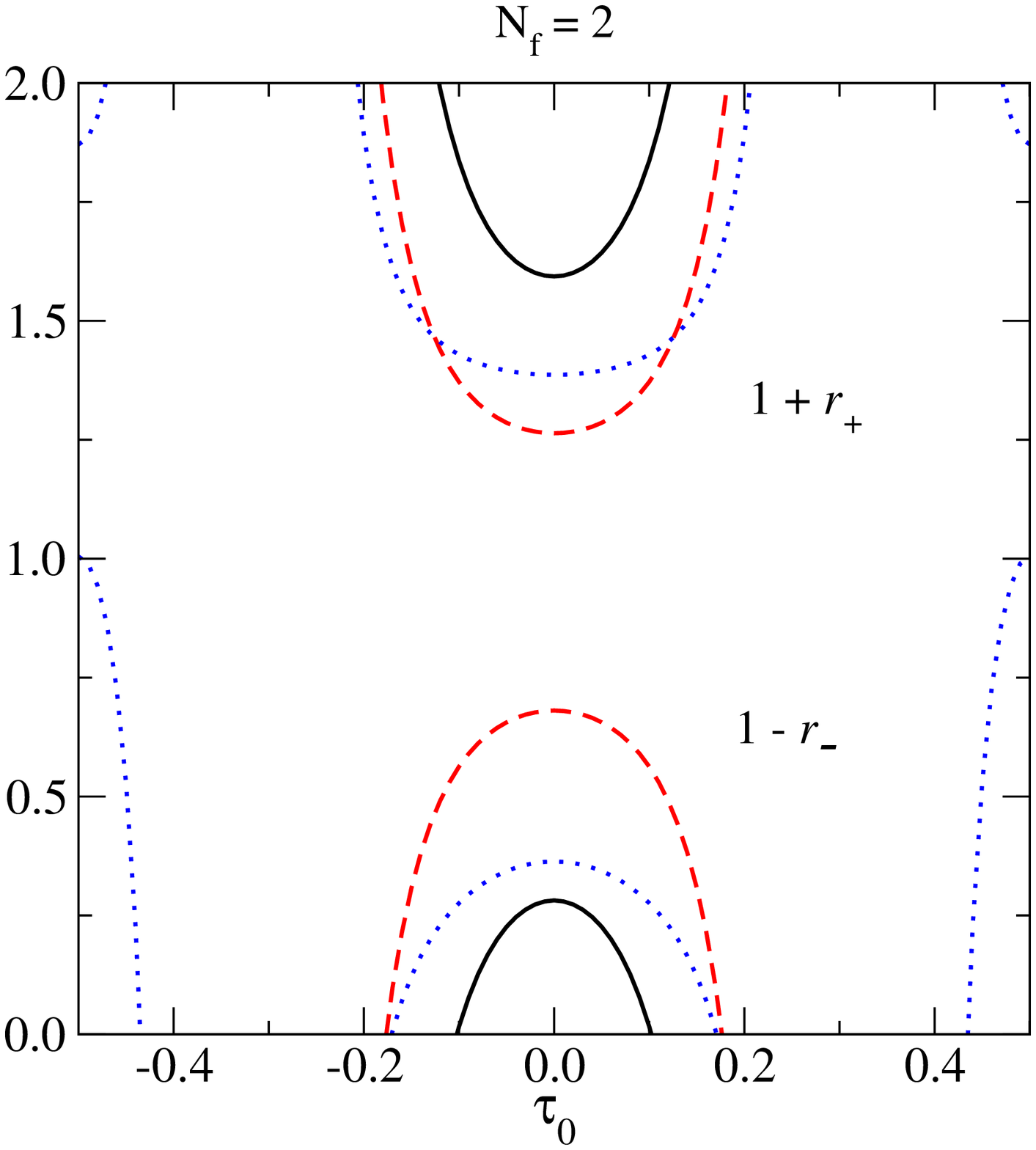}%
~~~\epsfysize=7.5cm\epsfbox{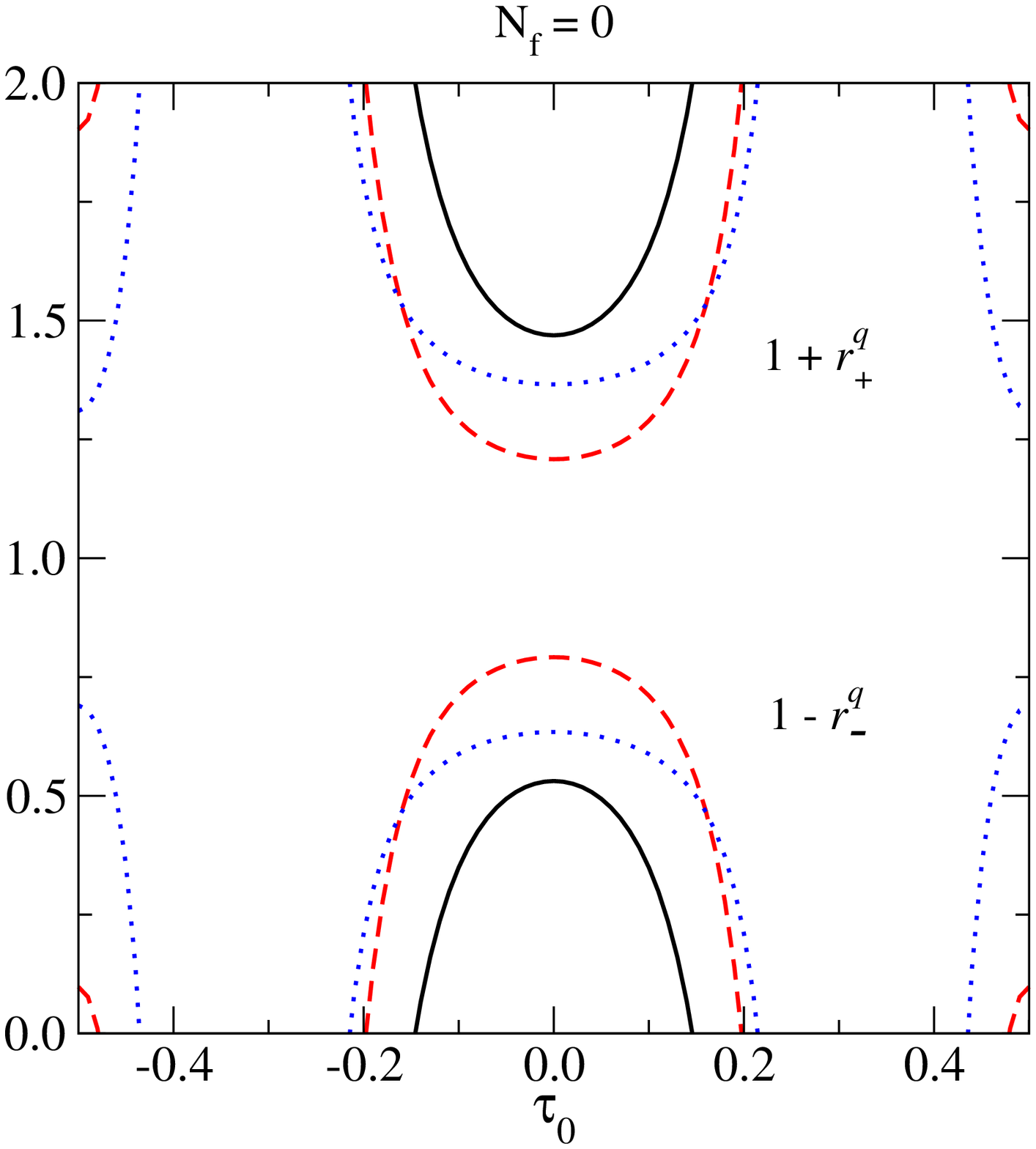}%

\caption{The functions $1\pm r_\pm$ (\eq\nr{R_r})
for $\Nf=2$ (left)  and $\Nf=0$ (right) as a function of 
$\tau_0\equiv z_0/T$, for $x_0=T/3$, $y_0 = 2 T/3$. 
The pion decay constant $F$ 
has been fixed to 93~MeV for $\Nf=2$ and to 110~MeV for $\Nf=0$. 
The solid line corresponds to $T=L=$2~fm, 
the dashed to $T=L=3$~fm,   
and the dotted to $T/2=L=$2~fm. \label{fig:rpmq}}
\end{center}
\end{figure}

The quenched result in the counting of ref.~\cite{zeromode} is
obtained by simply leaving out the term $1/\Nf$, because terms involving
$\alpha$ and $m_0^2$ are of higher order, like for the two-point 
correlator, and effects
from the coupling $K$ cancel at this order, as demonstrated in 
appendix~B (cf.\ \eqs\nr{Topo_q_first}, \nr{Topo_q_last}, 
\nr{Topo_zm_q_first}, \nr{Topo_zm_q_last}):
\be
 r_\pm^q(z_0) = 
 \frac{\rho}{(FL)^2}
 \Bigl[ 
  2 \beta_1 \rho^{-\fr32} 
 + 
  f_1(\tau_x) + f_1 (\tau_y) - h_1(\tau_x) - h_1(\tau_y) +
  H(\tau_x,\tau_y) 
 \Bigr]  
 \;. \la{rpm_quenched}
\ee
The most efficient way of evaluating numerically the amplitudes 
$\bar A_\nu$ and $\tilde A_\nu$ is by fixing the temporal position 
of the sources, $x_0$ and $y_0$, so that the three-point correlator
is measured as a function of the temporal position of the weak operator, 
$z_0$. In order to maximize the separation between the three operators 
we take $x_0  \sim T/3$ and $y_0 \sim  T-x_0$. 
The signal should be best when the weak operator is near the origin,  
and simultaneously the NLO correction $r_\pm(z_0)$ is minimized. 
The corresponding values of $(FL)^2 r^q_\pm(0)$ for various box shapes are 
given in Table~\ref{tab:rpmq}. There is a very strong dependence on $z_0$ 
(cf.\ \fig\ref{fig:rpmq}) 
which results from the fact that the functions $f_1(\tau_x)$ 
and $f_1(\tau_y)$ are divergent at $\tau_x=0$ and $\tau_y=0$, 
respectively, while they fall off exponentially, 
$\sim \exp(- 2\pi \rho \tau_x)$, $\sim \exp(- 2\pi \rho \tau_y)$,
away from these points\footnote{%
  These terms could 
  be decreased by increasing $\tau_x$, $\tau_y$ towards $1/2$, since 
  nothing dramatic happens as $\tau_x\to \tau_y$
  according to NLO expressions (cf.\ Table~\ref{tab:rpmq}). 
  However, this situation may be specific to the SU(4) case, 
  where there are no quark propagators connecting the pseudoscalar
  densities, while in the physical SU(3) case
  it is probably reasonable to keep 
  the pseudoscalar densities somewhat apart from other.}. 
Thereby a way to decrease these corrections is to increase $\rho = T/L$ \footnote{Note however that 
increasing $\rho$ takes us closer to the so-called $\delta$-regime~\cite{delta}.}.
In this respect the situation is opposite to that in 
ref.~\cite{prl} where left-handed currents appear 
in place of pseudoscalar densities; then
the NLO corrections increase rapidly with $T/L \gsim 2$~\cite{strategy}. 

%

%
\begin{table}[!t]
\small
\begin{center}
\begin{tabular}{ccccccc}
\hline\\[-2.0ex]
 $V$ & $x_0/a$ & $y_0/a$ & $\tau_x$ & $\tau_y$ & $(FL)^2 r^q_\pm(0)$ \\[1.5ex]
\hline\\[-2.0ex]
 $16^4$ & 4 & 12 &  0.25000 & 0.75000 & 0.89309 \\
 $16^4$ & 5 & 11 &  0.31250 & 0.68750 & 0.62839 \\
 $16^4$ & 6 & 10 &  0.37500 & 0.62500 & 0.51281 \\
 $16^4$ & 7 & 9  &  0.43750 & 0.56250 & 0.41875 \\
 $16^3 32$ & 11 & 21 & 0.34375 & 0.65625 & 0.45251 \\
 $16^3 32$ & 12 & 20 & 0.37500 & 0.62500 & 0.43174 \\
 $16^3 32$ & 13 & 19 & 0.40625 & 0.59375 & 0.38822 \\
 $16^3 32$ & 14 & 18 & 0.43750 & 0.56250 & 0.32157 \\
 $16^3 32$ & 15 & 17 & 0.46875 & 0.53125 & 0.23162 \\
 $24^4$ & 8  & 16 &  0.33333 & 0.66667 & 0.58270 \\
 $24^4$ & 9  & 15 &  0.37500 & 0.62500 & 0.51281 \\
 $24^4$ & 10 & 14 &  0.41667 & 0.58333 & 0.45099 \\
 $24^4$ & 11 & 13 &  0.45833 & 0.54167 & 0.38435 \\
\hline
\end{tabular}
\end{center}
\caption{Examples of values of $(F L)^2 r^q_\pm(0)$ for various box
shapes. Note that in the quenched limit, $r^q_\pm(0)$ is independent 
of the channel $\pm$, cf.\ \eq\nr{rpm_quenched}.}
\label{tab:rpmq}
\end{table}
%

\fig\ref{fig:rpmq} shows the result for $1\pm r^q_\pm$
for various box volumes, for $x_0 = T/3$, $y_0 = 2 T/3$.
Unfortunately, NLO corrections seem 
to be rather large at $L=2$~fm.
  
%
%




%
\section{Numerical results } 
\label{sec:num}

We have carried out a  numerical test of the method outlined above,
in the quenched approximation. 
Table~\ref{tab:simul} shows the simulation parameters. 
We have considered a symmetric lattice $T=L\simeq 2~$fm at two different 
lattice spacings in order to test for scaling violations in these observables.

%
\begin{table}[!t]
\vspace*{0.5cm}
\small
\begin{center}
\begin{tabular}{cccccccccc}
\hline 
 & $\beta$ &  $r_0/a$& $V$ & $N_\rmi{low}$ & $|\nu|$ & $N_\rmi{conf}^{|\nu|}$ & $\frac{x_0}{a},\frac{y_0}{a}$ & $Z_A$ & $am$\\[1.5ex]
\hline 
A$_1$ & 5.8458 & 4.026 & $16^4$ & 20 & 1-5  &  118,94   & 5,11 & 1.710 &  0.0015,0.0025, \\
   &        &       &        &    &  &  99,73,65 &      &       &  0.005 \\ \hline
A$_2$ & 6.0735 & 6.072 & $24^4$ &  20 & 2-5  &  92,63  & 8,16 & 1.505 & 0.002, 0.0033, \\
   &        &       &        &     &   &  51,55  &      &       & 0.0067\\
\hline
\end{tabular}
\end{center}
\caption{Parameters of the simulations.}
\label{tab:simul}
\end{table}
%

To evaluate the correlation functions of 
eqs.~(\ref{aatildep})--(\ref{bm}), we have computed chiral quark 
propagators on quenched background gauge configurations, using the 
Neuberger-Dirac operator with~$s=0.4$. For all details of the numerical 
implementation we refer the reader to refs.~\cite{methods,current,strategy}, 
whose techniques we adopt. 

Before presenting the actual data, 
let us describe how low-mode averaging (LMA) 
can be used to reduce statistical fluctuations in the signal.

\subsection{Low-mode averaging}
\la{se:lma} 

Our observables, eqs.~(\ref{aatildep})--(\ref{bm}), involve both
zero-mode wave functions and actual quark propagators. 
Since we employ low-mode preconditioning \cite{methods} for determining
the quark propagators, which requires the computation of a few lowest 
eigenvectors of the Dirac operator, we can use these in order to perform 
an extra averaging over time translations of the low-mode contribution 
to eqs.~(\ref{aatildep})--(\ref{bm}). In other words, the 
LMA technique~\cite{current,dgs} gives only an extra averaging 
over time in our case, but is nevertheless helpful as we will see, 
since the numerical overhead involved is negligible.

The main idea of LMA is to substitute the chiral propagator
 in eqs.~(\ref{aatildep})--(\ref{bm}) by 
\begin{eqnarray}
 P_{-\chi} S_m(z,x) P_\chi = 
 \sum_{k=1}^{N_\rmi{low}} \Psi_k(z) \otimes \Psi^\dagger_k(x) 
 + P_{-\chi} S^\rmi{sub}_m(z,x) P_{\chi}
 \;,
\end{eqnarray}
where $S^\rmi{sub}_m(z,x)$ is the inverse of the massive Dirac operator 
in the subspace orthogonal to the eigenspace of the approximate low-modes, and the chiral components of $\Psi_k$ are given by
\begin{eqnarray}
 P_{-\chi} \Psi_k(x) = w_k(x) , \;\;\;\; P_{\chi} \Psi_k(x)
 = \frac{1}{\alpha_k} P_{\chi} \gamma_5 D P_{-\chi} w_k(x)
 \;,
\end{eqnarray}
where $w_k(x)$ are the approximate eigenfunctions of the operator 
$D^\dagger D$ for the eigenvalue $\lambda_k$, while 
$\alpha_k = \sqrt{\lambda_k}$.  We indicate how this works with 
the two-point correlator. 

The LMA evaluation of eq.~(\ref{bp}) is based on the separation
\begin{eqnarray}
B_\nu &=& B_\nu^l + B_\nu^h ,
\end{eqnarray}
where the ``high-mode part'' reads
\begin{eqnarray}
B^h_\nu(x_0-z_0) &=& \lim_{m\rightarrow 0}~\frac{1}{L^3} \int_{\vec{z}}
\Bigl\langle \sum_{i=1}^{|\nu|}
v_i^\dagger(z) \gamma_0 {\tilde\eta}_i(z;x_0) 
\Bigr\rangle_\nu \, ,
\end{eqnarray}
with 
\begin{eqnarray}
 {\tilde\eta}_i(z;x_0) \equiv  \partial_{x_0}
 \int_{\vec{x}}  P_{-\chi} S^\rmi{sub}_m(z,x) P_\chi v_i(x) \;,
\end{eqnarray}
while the low-mode contribution is 
\begin{eqnarray}
 B^l_\nu(t) &=& \lim_{m\rightarrow 0}~\sum_{i=1}^{|\nu|}\sum_{k=1}^{N_\rmi{low}}
 \frac{1}{V} \int_{x,z} \delta(x_0 - z_0 - t)
 \left\langle
 v_i^\dagger(z) \gamma_0 P_- \Psi_k(z) \partial_{x_0} 
 \left[ \Psi^\dagger_k(x) P_+ v_i(x)\right]
 \right\rangle_\nu 
 \;. \nn 
\end{eqnarray}
The LMA evaluation of the three-point function is
carried out in complete analogy.

\subsection{Two-point function}

\begin{figure}[tb]
\begin{center}
\includegraphics[width=7.5cm]{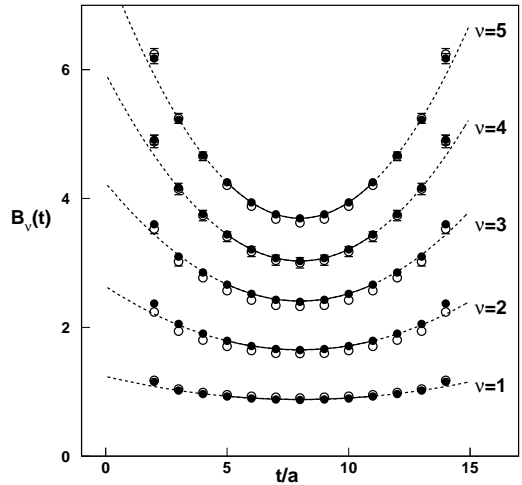}
\caption{The measured $T B_\nu(t)$ for lattice A$_1$ and quark mass
$am = 0.0015$. Open/full symbols correspond to data without/with LMA. 
Error bars are in most cases smaller than the symbol sizes.}
\label{fig:B}
\end{center}
\end{figure}

In \fig\ref{fig:B} we show results for
the two-point correlator $B_\nu(t)$ (cf.\ \eqs\nr{bp}, \nr{bm})
in  different topological sectors
at the lightest quark mass. The open/full symbols corresponds 
to the results without/with LMA. 
There is a strong dependence on $|\nu|$, as expected from $\chi$PT. 

%
%

Since there is clear evidence for NLO corrections, 
we consider a two-parameter fit of the form
\begin{equation}
 T B_\nu(t) =  \alpha_\nu + 2 \beta_\nu h_1\left(\tau\right)
 \;, \quad \tau = {t \over T}
 \;,
 \label{eq:fit}
\end{equation}
where $h_1$ is from \eq\nr{h1}.
The temporal dependence in all sectors is perfectly compatible with the function $h_1(\tau)$ as illustrated by the solid lines in \fig\ref{fig:B}, which are the results of the fits in the time interval $\Delta t=5a-11a$.
The small dependence on the quark mass is perfectly linear, so the values of $\alpha_\nu$ and $\beta_\nu$ are linearly extrapolated to the zero mass limit.  The results for $\alpha_\nu$ and $\beta_\nu$ in the chiral limit are summarized in Table~\ref{tab:ab}. The jackknife procedure has been used to estimate the errors.  

\begin{figure}[tbp]
\begin{center}
\includegraphics[width=7.5cm]{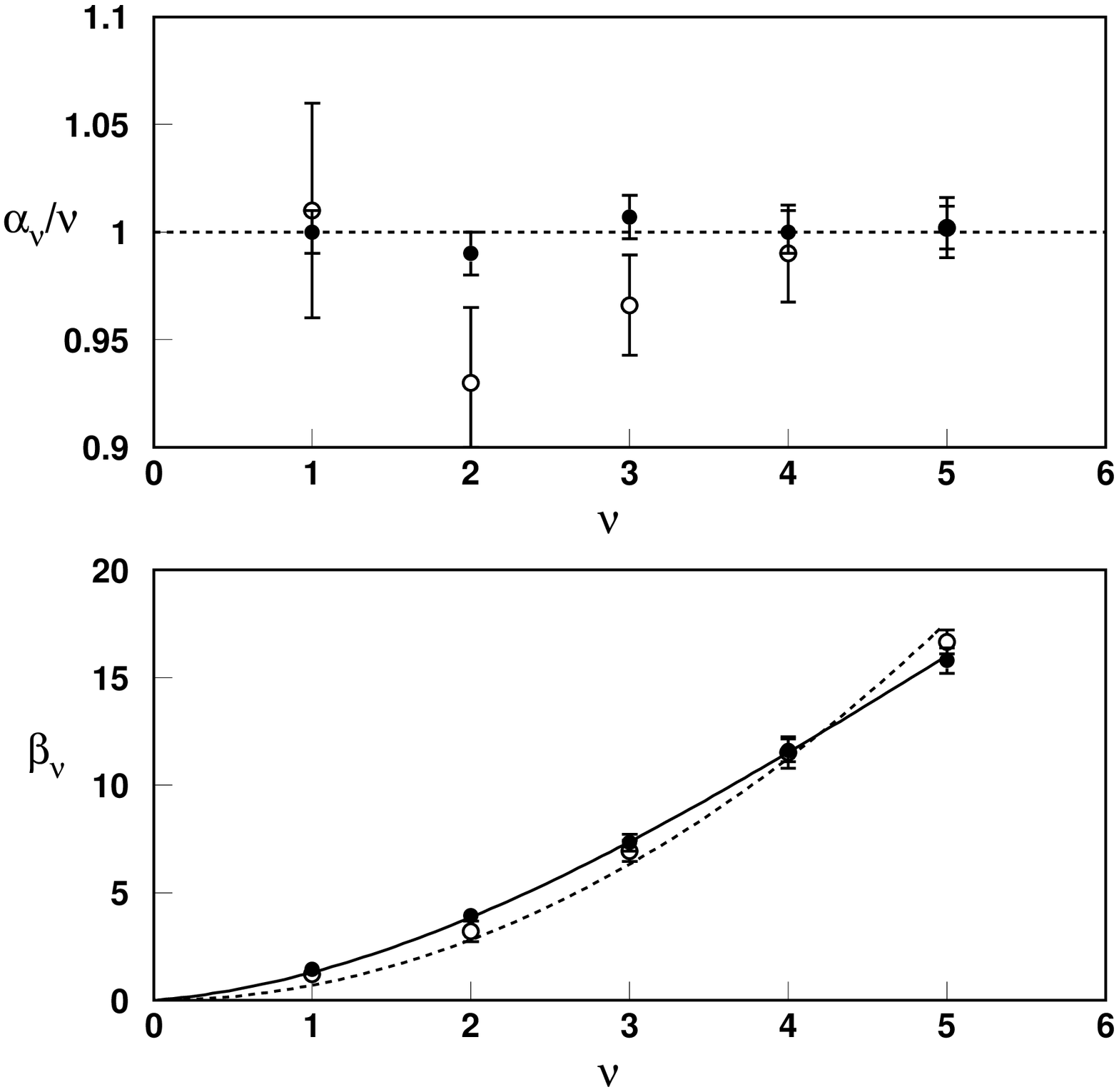}%
\includegraphics[width=7.5cm]{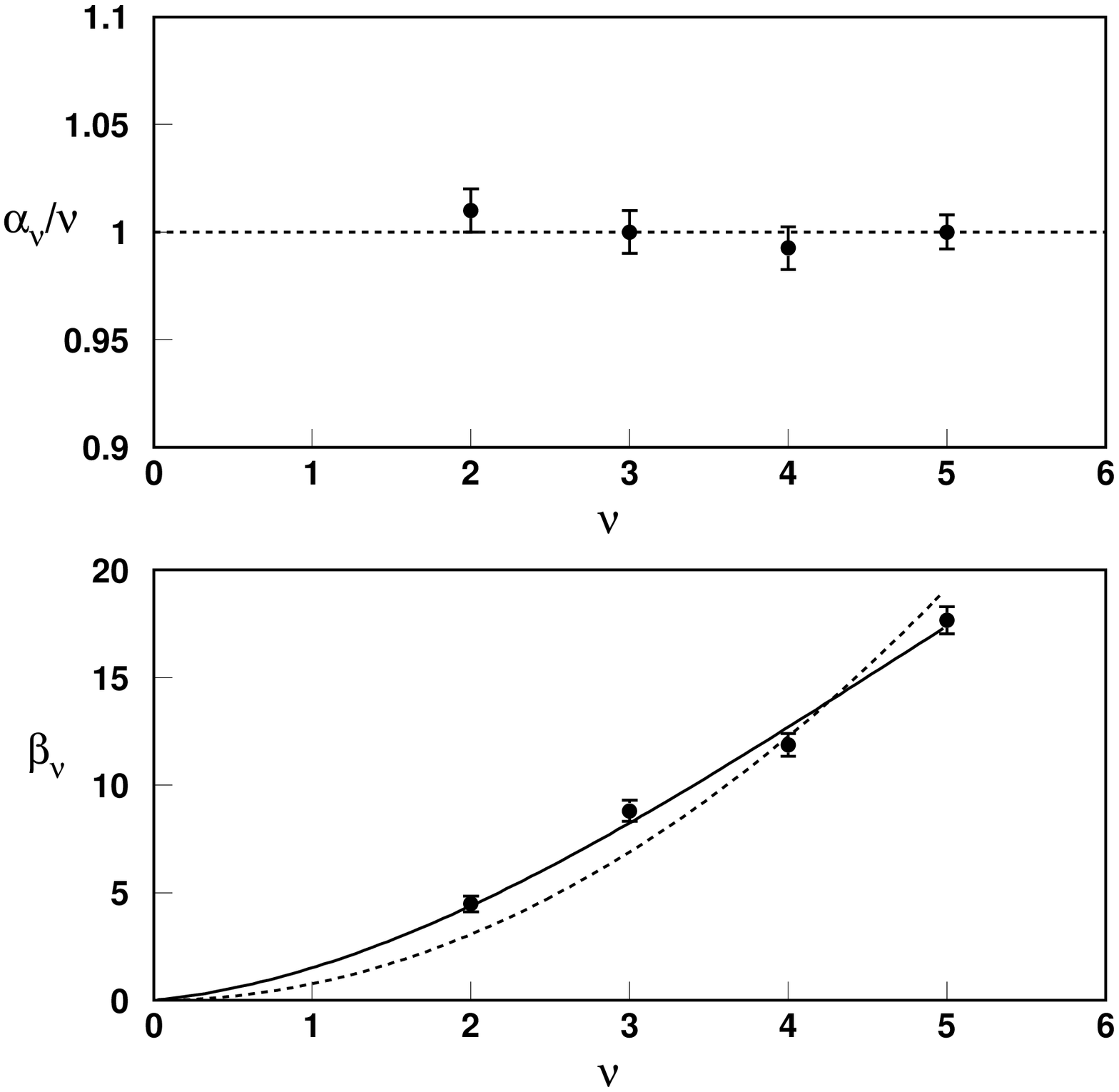}
\caption{Top: $\alpha_\nu/|\nu|$  versus $|\nu|$ 
for lattice A$_1$ (left) and lattice A$_2$ (right). 
The dashed line is the NLO expectation. Bottom: $\beta_\nu$ 
versus $|\nu|$ for the same lattices. 
The dashed line is the best fit NLO prediction, 
the solid line is the best fit NNLO prediction. In both cases 
open/full symbols correspond to without/with LMA. }
\label{fig:nudep_A1}
\end{center}
\end{figure}

In \fig\ref{fig:nudep_A1} 
we show the results 
for $\alpha_\nu$ and $\beta_\nu$ as a function of $|\nu|$, together 
with the NLO expectations. In the case of $\alpha_\nu$, 
the NLO prediction $\alpha_\nu = |\nu|$ is extremely well reproduced 
at the per cent level.  The prediction for $\beta_\nu$, on the other hand, 
depends on $F$. The dashed line in the figure corresponds 
to a fit to the NLO prediction leaving 
$F$ as a free parameter.  The best fit values are 
$(F L)_\rmi{A$_1$} =1.19(2)$ and $(F L)_\rmi{A$_2$}= 1.14(2)$  
with $\chi^2/\mbox{d.o.f} \sim 9$ and 12, respectively, 
which are rather bad. 
Clearly the $|\nu|$ dependence is not properly reproduced at NLO, however 
it seems that the discrepancy could be ascribed 
to higher order chiral corrections.

We have seen that the Ward identity, \eq\nr{wi}, 
relates $B_\nu(t)$ to the topological zero-mode contribution 
in the correlator
of two pseudoscalar densities. This quantity was studied 
in ref.~\cite{zeromode}, 
and actually the chiral corrections were computed 
to one order higher than here. Although the expressions are rather 
complicated and involve new time-dependent functions, a convenient way 
to try to include these corrections is to consider a Taylor expansion 
around the middle point, $t= T/2$. From the results of 
ref.~\cite{zeromode}, we expect that
\begin{equation}
  T B_\nu(t) =  \gamma_\nu + 
 \delta_\nu \left({t\over T} - {1 \over 2}\right)^2 + ... 
 \;, 
 \label{fit:par}
\end{equation}
where\footnote{We have set $\alpha=0$, since this parameter was consistent
with zero in the analysis of ref.~\cite{zeromode}.} 
\begin{equation}
 \delta_\nu = {\rho \over (F L)^2}~\left\{ \nu^2 + 
 {\rho |\nu| \over (FL)^2 }  \left[  
 - \beta_1 \rho^{-\fr32} - {1 \over 24} \left( {7\over 3} + 2 \nu^2 
 -2 \langle \nu^2 \rangle\right) + {\gamma_1\over 2}\right]
 \right\}
 \;.
 \label{eq:delta}
\end{equation}
For a symmetric box with $T=L$ 
the shape coefficients read $\rho =1$, $\beta_1= 0.14046098$,  
and $\gamma_1 = -0.05712765$. 

Numerically the results for $\delta_\nu$ in a fit of the form 
in eq.~(\ref{fit:par}) are identical  to those for $\beta_\nu$ in 
Table~\ref{tab:ab}.  The solid line in the lower plot of 
Fig.~\ref{fig:nudep_A1} corresponds to the prediction of eq.~(\ref{eq:delta}) 
using $\langle \nu^2 \rangle = \chi_t V = 0.059 V/r_0^4$ \cite{topsus} 
and leaving $F$ as a free parameter. The result of the fit 
is $(F L)_\rmi{A$_1$}= 1.12(2)$ and $(FL)_\rmi{A$_2$}=1.07(2)$ with 
$\chi^2/\mbox{d.o.f} = 0.2,1.3$, respectively. The results for $F$ 
are essentially the same as the central values 
in ref.~\cite{zeromode} (this is almost trivial 
in the light of \se\ref{ss:wi} below)\footnote{Note that 
a similar physical box size was used  in ref.~\cite{zeromode} as in the present study.}; however, 
the current error bars are much smaller, 
because higher topological sectors which are less noisy have been considered.
Besides, we have not 
assigned any errors to $\alpha$ and $\chi_t$, 
as we did in \cite{zeromode}. The results  
are in reasonable agreement also with the determination of $F$ from 
the left-current two-point correlator in ref.~\cite{current}, although
it must be noted that the box size was significantly smaller there.

%
\begin{table}[!t]
\small
\begin{center}
\begin{tabular}{ccccc}
\hline\\[-2.0ex]
$|\nu|$ & $\alpha_\nu$(A$_1$) & $\beta_\nu$ (A$_1$) & $\alpha_\nu$(A$_2$) & $\beta_\nu$ (A$_2$)   \\[1.5ex]
\hline\\[-2.0ex]
$1$ & $1.00(1)$ &  $1.4(3)$ &  & \\[1.5ex]
$2$ & $1.98(2)$ &  $3.9(4)$ & 2.02(2) & 4.5(4)\\[1.5ex]
$3$ & $3.02(3)$ & $7.3(4)$  & 3.00(3) & 8.8(5)\\[1.5ex]
$4$ & $4.00(4)$ &  $11.6(5)$ & 3.97(4) & 11.9(5) \\[1.5ex]
$5$  & $5.01(5)$ & $15.8(6)$  & 5.00(4) & 17.7(6) \\[1.5ex]
\hline
\end{tabular}
\end{center}
\caption{Results for $\alpha_\nu$ and $\beta_\nu$ from  
a fit to eq.~(\ref{eq:fit}) of 
the measured two-point correlator, subsequently extrapolated 
to the zero-mass limit.}
\label{tab:ab}
\end{table}
%


%
%


\subsection{Ward identity and chiral extrapolation}
\la{ss:wi}

The Ward identity of eq.~(\ref{wi}) is a good test of the
extrapolation $m\to 0$ needed  for the correlator $B_\nu(t)$.

\begin{figure}[tb]
\begin{center}

\vspace{1cm}
\includegraphics[width=9cm]{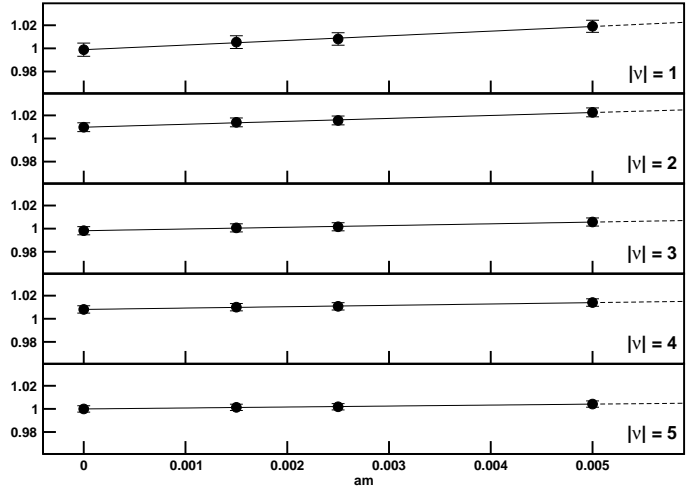}


\caption{Chiral extrapolation of the time-averaged ratio $D_\nu(t)/B_\nu(t)$
(cf.\ \eq\nr{wi_expl}) normalized to $\hat{Z}_A$ from ref.~\cite{hj}, for lattice A$_1$.}
\label{fig:za}
\end{center}
\end{figure}

In \fig\ref{fig:za} we show results 
for the ratio $Z_A \equiv D_\nu(t)/B_\nu(t)$ (notation as in \eq\nr{wi_expl})
as a function of $a m$ for the different topological sectors, 
normalized to the value ${\hat Z}_A$ obtained by conventional means 
in ref.~\cite{hj}. In the limit $m\rightarrow 0$, the ratio should 
approach unity in all topological sectors. The level of agreement between 
the different sectors and with ${\hat Z}_A$ 
is shown in Table~\ref{tab:wi}.
%
\begin{table}[!t]
\small
\begin{center}
\begin{tabular}{ccc}
\hline\\[-2.0ex]
$|\nu|$ & $Z_A/\hat {Z}_A$(A$_1$) & $Z_A/\hat {Z}_A$(A$_2$) \\[1.5ex]
\hline\\[-2.0ex]
$1$ &  $1.000(6)$ & \\[1.5ex]
$2$ &  $1.009(4)$ & 0.991(3) \\[1.5ex]
$3$  & $0.999(4)$  & 0.995(3) \\[1.0ex]
$4$  &  $1.008(3)$ & 0.999(3)\\[1.5ex]
$5$  & $1.000(3)$  & 1.002(3) \\[1.0ex]
\hline
\end{tabular}
\end{center}
\caption{The ratio $Z_A \equiv D_\nu/B_\nu$ 
(cf.\ \eq\nr{wi_expl}), obtained from the saturation with zero modes of 
the Ward Identity (with LMA treatment of the non-zero mode part of $B_\nu$), 
normalized to the conventionally determined $\hat Z_A$~\cite{hj}. 
The errors do not include the error on 
${\hat Z}_A$ which is about 3 per mille.}
\label{tab:wi}
\end{table}
Given that only the zero-mode contributions to both sides of the 
Ward Identity are included, this is a strong check of the whole procedure, 
and furthermore indicates that the small residual extrapolation 
to zero quark mass is under good control. 

%
%

\subsection{Three-point function}

\begin{figure}[tb]
\begin{center}
\includegraphics[width=9cm]{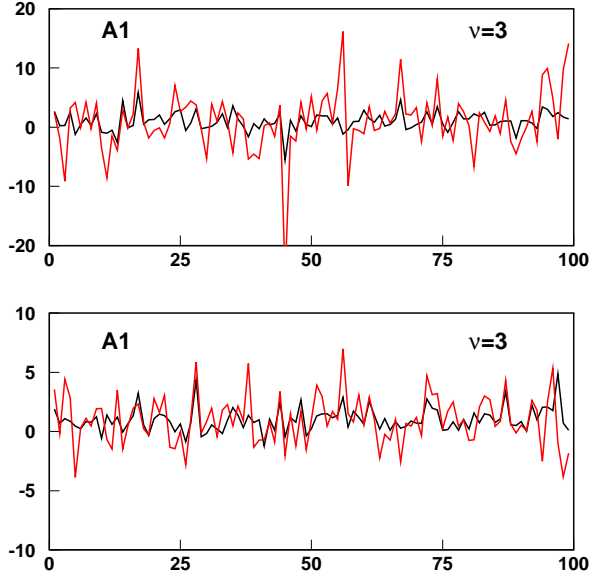}
\caption{The Monte Carlo history of the three-point 
correlators  $\bar A_\nu - \tilde A_\nu$ (top) and 
$\bar A_\nu + \tilde A_\nu$ (bottom), 
at $z_0=0$, normalized to the average, on the lattice A$_1$, 
for $|\nu|=3$ and $am=0.0015$. The dark line corresponds 
to the LMA and the light one to the non-LMA amplitude.}
\label{fig:lmavsnonlma}
\end{center}
\end{figure}

In \fig~\ref{fig:lmavsnonlma} we show one example of 
a Monte Carlo history for the three-point functions
$\bar A_\nu \pm \tilde A_\nu$,  
\eqs\nr{aatildep}, \nr{aatildem}, 
with and without LMA. Clearly LMA improves 
the signal significantly. The improvement is more pronounced for 
the smaller topologies and masses as expected.  In all of 
the following we consider only the LMA results.

%
%

\begin{figure}[tb]
\begin{center}
\includegraphics[width=12cm]{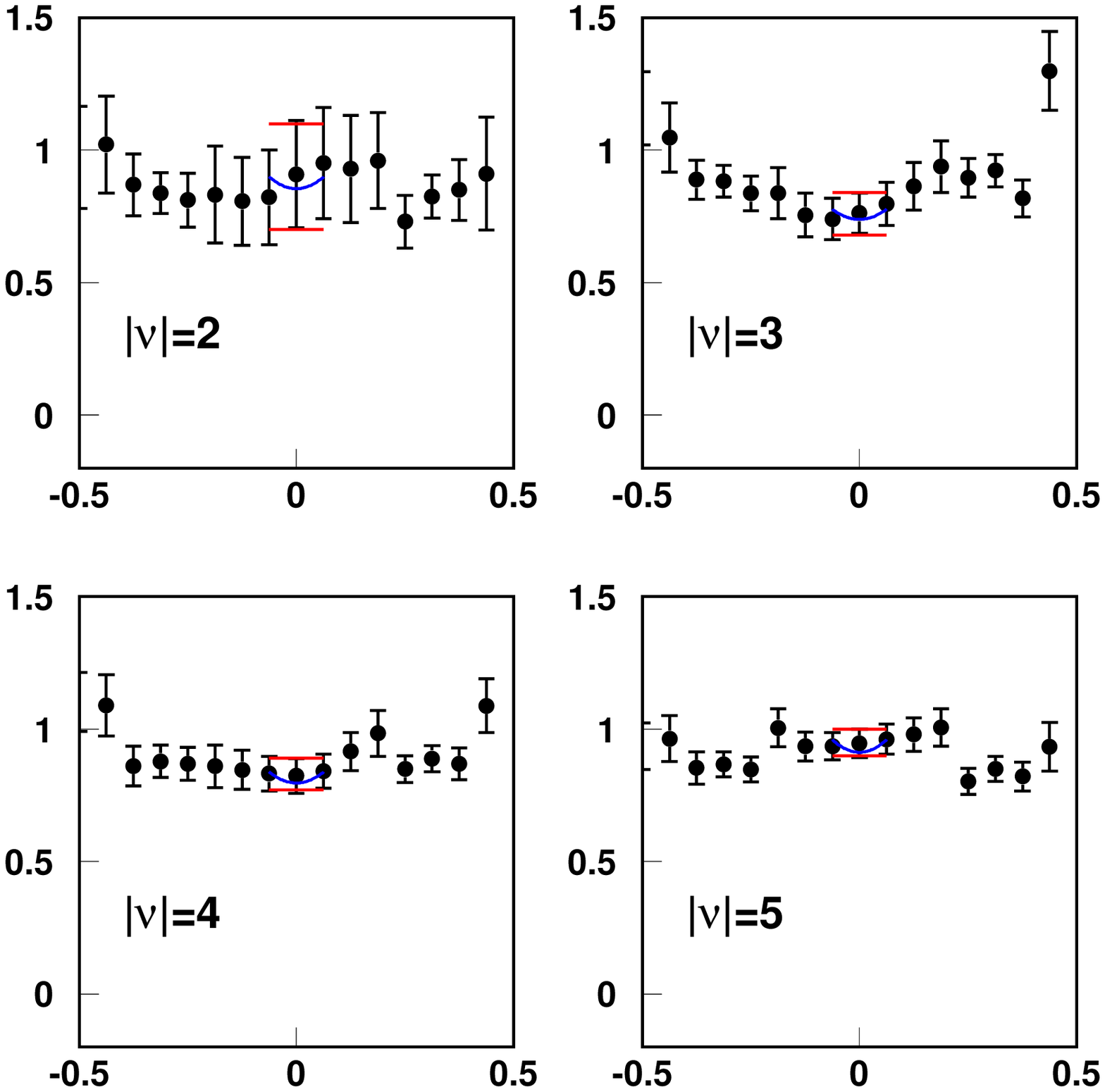}
\caption{$R_\nu^+/(1-1/|\nu|)$ 
in different topological sectors
for the lattice A$_1$ as 
a function of $z_0/T$ for fixed $x_0 = 5 a$, $y_0 = 11 a$,  
at the smallest quark mass $a m =0.0015$. 
The horizontal lines represent the 1$\sigma$ boundaries 
of a LO fit, while the curved line is the best NLO fit. }
\label{fig:Rp}
\end{center}
\end{figure}

\begin{figure}[tb]
\begin{center}
\includegraphics[width=12cm]{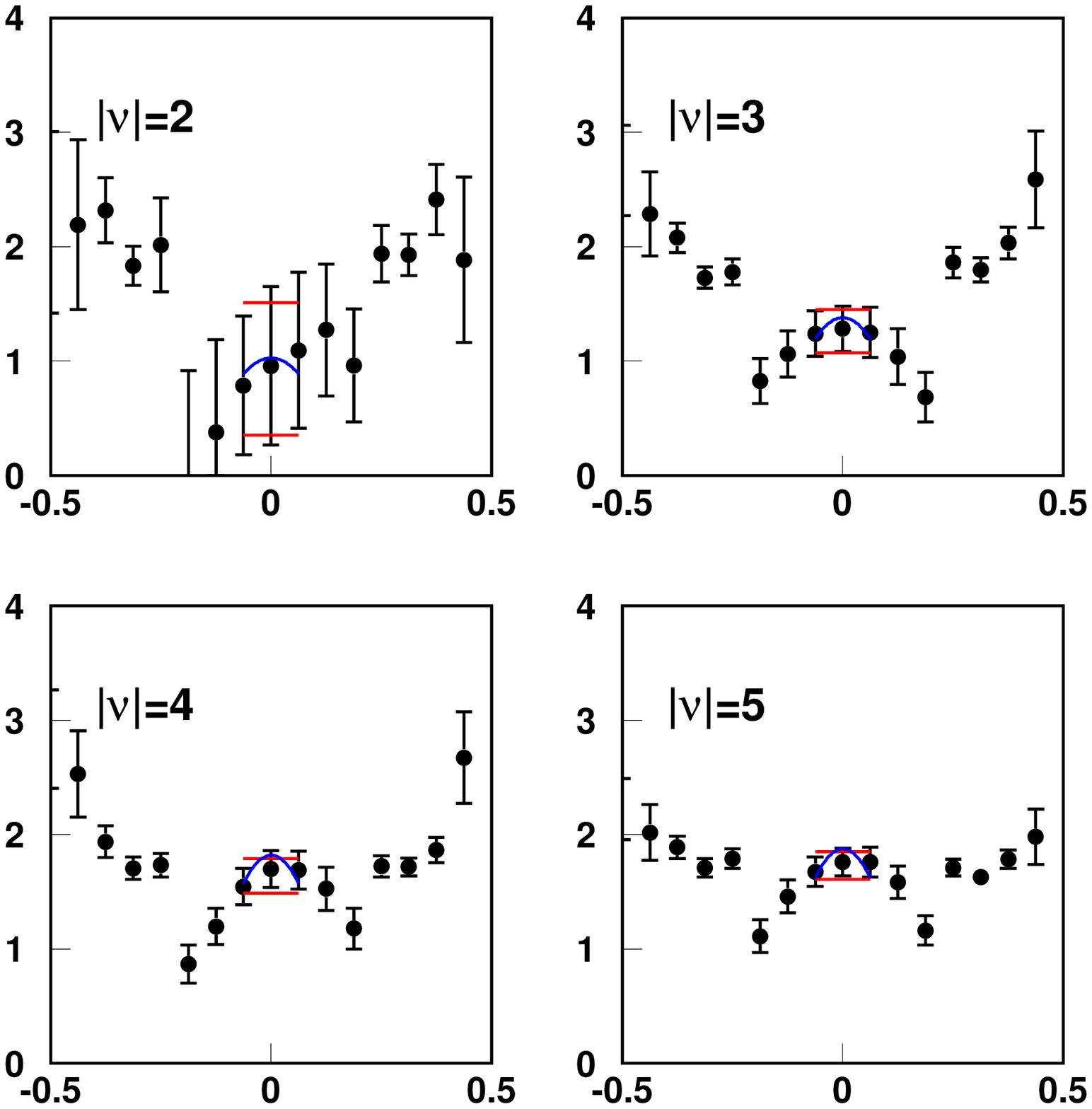}
\caption{$R_\nu^-/(1+1/|\nu|)$
in different topological sectors
for the lattice A$_1$ as 
a function of $z_0/T$ for fixed $x_0 = 5 a$, $y_0 = 11 a$,  
at the smallest quark mass $a m =0.0015$. 
The horizontal lines represent the 1$\sigma$ boundaries 
of a LO fit, while the curved line is the best NLO fit. }
\label{fig:Rm}
\end{center}
\end{figure}

In \figs\ref{fig:Rp}, \ref{fig:Rm} we show the results for 
the bare ratios $R_\nu^\pm(x_0-z_0,y_0-z_0)/(1\mp1/|\nu|)$ on lattice A$_1$ 
as a function of $\tau=z_0/T$, at fixed $x_0=5a$, $y_0=11 a$. 
The quark mass is $am=0.0015$. There is a clear signal near $\tau=0$. 
However, the temporal dependence does not seem to be as 
pronounced as expected from NLO $\chi$PT. The pattern is  
similar for the lattice A$_2$.

%
%

%
%



\begin{figure}[tb]
\begin{center}
\includegraphics[width=12cm]{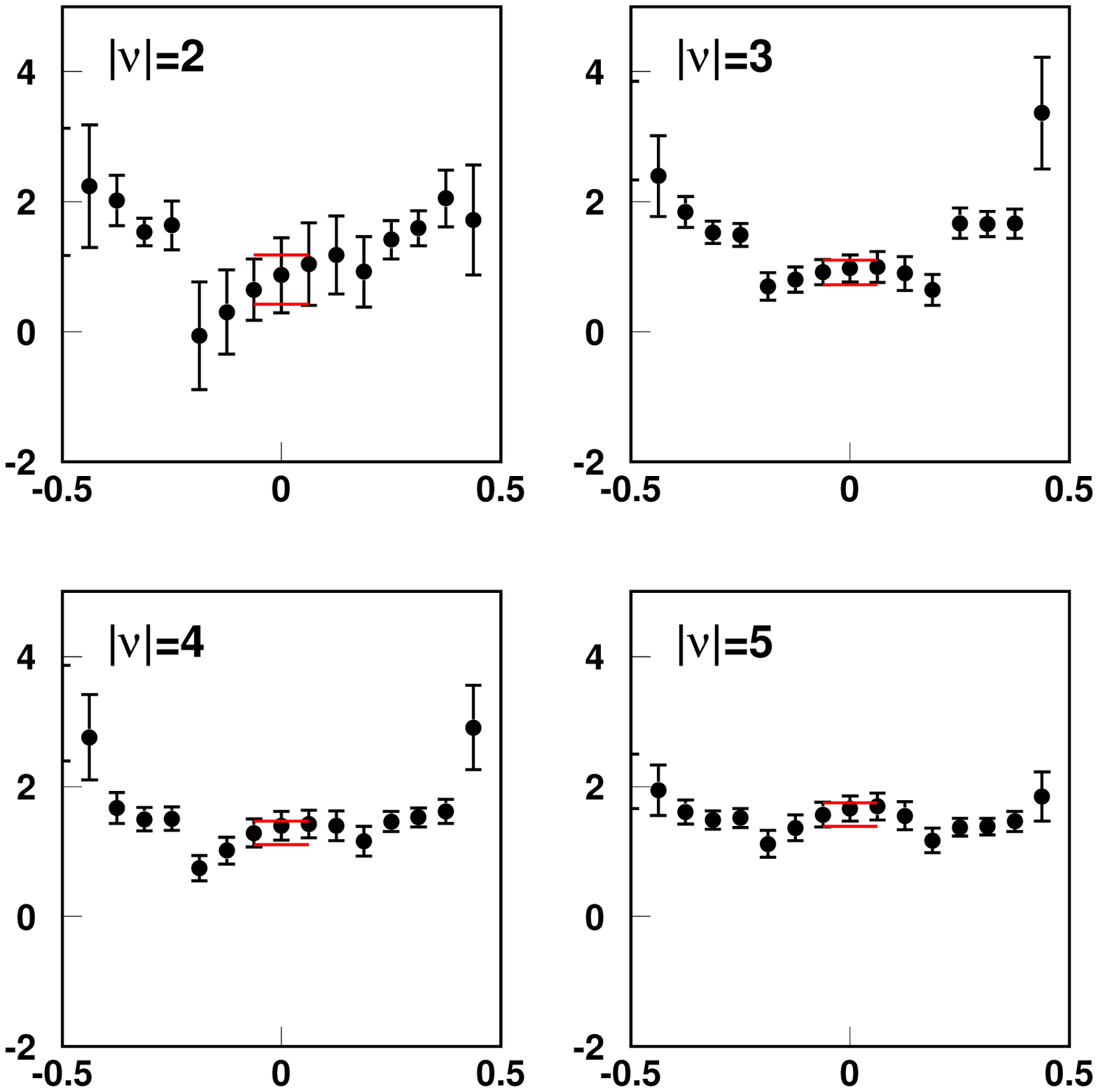}
\caption{$R_\nu^+ R_\nu^-/(1-1/\nu^{2})$
in different topological sectors 
on the lattice A$_1$ as a function of $z_0/T$ 
for fixed $x_0 = 5 a$, $y_0 = 11 a$,  
at the smallest quark mass $am =0.0015$. 
The bands represent the 1$\sigma$ fits to a constant.}
\label{fig:prod_16}
\end{center}
\end{figure}

An interesting combination to consider is the product $R_\nu^+
R_\nu^-$, since NLO corrections cancel in this quantity
(cf.\ \eq\nr{R_r}). 
Writing the weak LECs as
\be
 g_1^\pm = [g_1^\pm]^\rmi{bare}\; {k_1^\pm Z_{11}^\pm \over Z_A^2} 
 \;, \la{gpm_bare}
\ee
\eqs\nr{R_r} and \nr{match} imply that we may expect:
\begin{equation}
  R_\nu^+ R_\nu^- = [g_1^+ g_1^-]^\rmi{bare} 
 \left(1-{1\over |\nu|^2} \right) + \ldots \;.
\end{equation}
The results for the lattice A$_1$ are shown in \fig\ref{fig:prod_16}.
A constant fit 
around $z_0= 0$, followed by a chiral extrapolation in each topological 
sector, gives the values shown in Table~\ref{tab:prod}. 
The agreement with the result obtained from the left-current 
three-point functions \cite{prl} is quite good. Note that the renormalization factors for the lattice A$_1$ are the same as those in ref.~\cite{prl} and therefore the bare couplings can be compared directly. Those for the lattice A$_2$ have not been evaluated. However, the difference is expected to be well below the statistical uncertainty. Indeed a LO computation in bare perturbation theory yields variations at the 1--2$\%$ level, and the results of ref.~\cite{renorm} indicate that this perturbative estimate is realistic.

%
%

%
%


%
\begin{table}[!t]
\small
\begin{center}
\begin{tabular}{cccc}
\hline\\[-2.0ex]
$|\nu|$ & $[g_1^+ g_1^-]^\rmi{bare}$ (A$_1$)  & $[g_1^+ g_1^-]^\rmi{bare}$ (A$_2$) & $[g_1^+ g_1^-]^\rmi{bare}$ \cite{prl} \\[1.5ex]
\hline\\[-2.0ex]
$2$ & 0.73(53) &  0.72(39) & \\[1.5ex]
$3$  &  0.94(20) &  1.10(34) &\\[1.0ex]
$4$  &  1.37(20) & 1.60(36) & \\[1.5ex]
$5$  &  1.64(20) &  1.50(17) &\\[1.0ex]
\hline\\[-2.0ex]
  w.a. ($|\nu| > 2$)  &   1.32(11) & 1.45(14) & 1.47(12) \\[1.0ex]
  $\chi^2/{\rm d.o.f.}$  &   3.1 & 0.7  &  \\[1.0ex]
\hline
\end{tabular}
\end{center}
\caption{Values of the product of the bare couplings obtained from fits 
to a constant in the time interval $|z_0|\leq a$ for lattice A$_1$ and 
$|z_0|\leq 2 a$ for A$_2$. The last row shows the weighted averages (w.a)
over topological sectors $|\nu| > 2$ for the two lattices and also 
the result obtained in ref.~\cite{prl} with a different method.}
\label{tab:prod}
\end{table}
%

In order to then estimate $[g_1^+]^\rmi{bare}$ and
$[g_1^-]^\rmi{bare}$ individually, we consider the following 
fits of the LMA data:

\paragraph{Fit A.} At LO we expect (cf.\ \eq\nr{R_r})
\begin{eqnarray}
 R_\nu^\pm = [g_1^\pm]^\rmi{bare} \left(1\mp {1 \over |\nu|}\right)
 \;, 
\end{eqnarray}
therefore we fit  $R_\nu^\pm/\left(1\mp {1 \over |\nu|}\right)$
to a constant around $z_0=0$. The results are shown as 
the bands in \figs\ref{fig:Rp}, \ref{fig:Rm}.  

The results for the bare couplings are listed in Table~\ref{tab:lo}. 
The results on the  
lattices A$_1$ and A$_2$ are again perfectly compatible, 
which implies that scaling violations are well below
the statistical uncertaintities. 
There is a significant difference with the $[g_1^\pm]^\rmi{bare}$ 
obtained from a LO matching of left-current three-point functions 
in ref.~\cite{prl}: while our $[g_1^-]^\rmi{bare}$ is smaller, 
our $[g_1^+]^\rmi{bare}$ is larger.  This is, however, not unexpected, 
given that chiral corrections tend to decrease $[g_1^-]^\rmi{bare}$ 
and  increase $[g_1^+]^\rmi{bare}$ in ref.~\cite{prl}, 
and do the opposite in the present work.
%
\begin{table}[!t]
\small
\begin{center}
\begin{tabular}{ccccccc}
\hline\\[-2.0ex]
 & 
 \multicolumn{3}{c}{$[g_1^-]^\rmi{bare}$} & 
 \multicolumn{3}{c}{$[g_1^+]^\rmi{bare}$} \\[1.5ex]  
$|\nu|$ & 
 A$_1$ &
 A$_2$ &
 \cite{prl} &
 A$_1$ &
 A$_2$ & 
 \cite{prl}   \\[1.5ex]
\hline\\[-2.0ex]
$1$ & 
 3.1(1.6) &
  & 
  &
  & 
  & 
 \\[1.5ex]
$2$ &
 0.8(7) & 
 1.7(0.7)&
  & 
 0.9(2) &
 0.44(13) &
   \\[1.5ex]
$3$  & 
 1.2(2) &
 1.46(35) &
  & 
 0.76(8) &
 0.80(10) &
  \\[1.0ex]
$4$  & 
 1.65(15) &
 1.85(31) &
  & 
 0.83(6) &
 0.89(10) &
  \\[1.5ex]
$5$  &
  1.73(12) &
  1.74(15) &
   & 
  0.95(5) &
  0.86(6) &
  \\[1.0ex]
\hline\\[-2.0ex]
w.a. ($|\nu| > 2$)  &
  1.60(8) & 
  1.72(13) &
  2.42(13) &
  0.87(3) &
  0.85(5) &
  0.60(4)\\[1.0ex]
$\chi^2/{\rm d.o.f.}$  &
  2.6 & 
  0.4 &
    &
  2.4 &
  0.2 &
    \\[1.0ex]
\hline
\end{tabular}


\end{center}
\caption{Values of the bare couplings obtained from a LO fit
 in the time interval 
 $|z_0|\leq  a$ on lattice A$_1$ and $|z_0|\leq 2 a$  on lattice A$_2$.}
\label{tab:lo}
\end{table}
%

\paragraph{Fit B.} At NLO we expect (cf.\ \eq\nr{R_r})
\begin{eqnarray}
 R_\nu^\pm = [g_1^\pm]^\rmi{bare} 
 \left(1\mp {1 \over |\nu|}\right) \Bigl[ 1\pm r_\pm^q(z_0) \Bigr]
 \;. 
\end{eqnarray}
Numerical values of $(FL)^2 r_\pm^q(z_0)$ 
for small $z_0$ are shown in Table~\ref{tab:r}.  
%
\begin{table}[!t]
\small
\begin{center}
\begin{tabular}{ccccc}
\hline\\[-2.0ex]
$z_0$ & $[(FL)^2 r_\pm^q(z_0)]_\rmi{A$_1$}$ &
 $[(FL)^2 r_\pm^q(z_0)]_\rmi{A$_2$}$ \\[1.5ex]
\hline\\[-2.0ex]
    0 &    0.62839  & 0.58270 \\[1.5ex]
    $a$ &  0.73420  & 0.61454 \\[1.5ex]
   $2 a$ & 1.23214  & 0.72759 \\[1.0ex]
\hline
\end{tabular}
\end{center}
\caption{Values of $(FL)^2 r_\pm^q(z_0)$ for $z_0=0, a, 2a$, 
with $x_0=5 a$, $y_0 =11 a$ for lattice A$_1$ and 
$x_0=8 a$, $y_0=16 a$ for lattice A$_2$.}
\label{tab:r}
\end{table}
%
Taking a value $FL \sim 1.1$, the NLO prediction gives 
a stronger temporal dependence than that seen in the data. The fits have 
a bad $\chi^2$ if more that 3(5) points are included for lattice A$_1$(A$_2$).
This already indicates that NNLO could be significant. With so
few points it does not make sense to do a two parameter fit, leaving 
the normalization of the NLO correction free, because there is not 
much curvature in the data. Therefore we fix $(F L)^2=1.14-1.40$, 
corresponding to the range of values obtained from fitting various 
two-point functions here and in ref.~\cite{2ptnew}. We then perform 
linear one-parameter fits for $[g_1^\pm]^\rmi{bare}$. 
The results are summarized in Table~\ref{tab:nlo}.
%
\begin{table}[!t]
\small
\begin{center}
\begin{tabular}{cccc}
\hline\\[-2.0ex]
 & 
 \multicolumn{3}{c}{$[g_1^-]^\rmi{bare}$}
 \\[1.5ex]  
$|\nu|$ & 
 A$_1$ &
 A$_2$ &
 \cite{prl}
  \\[1.5ex]
\hline\\[-2.0ex]
$2$ &
 2(2) -- 1.6(1.4)  &
 4(2) -- 3(1) &
 \\[1.5ex]
$3$  &
 3.1(5) -- 2.5(4) &
 3.4(8) -- 2.7(7) &
 \\[1.0ex]
$4$  &
 4.2(4) -- 3.3(3) &
 4.3(7) -- 3.5(6)&
 \\[1.5ex]
$5$  & 
 4.4(3) -- 3.4(2) & 
 4.0(4) -- 3.2(3) &
 \\[1.0ex]
\hline\\[-2.0ex]
w.a. ($|\nu| > 2$) & 
 4.1(2) -- 3.2(2)  &
 4.0(3) -- 3.2(3) &
 2.33(11)
 \\
 $\chi^2 /{\rm d.o.f.}$  &
  2.5 -- 2.0  &
  0.4 -- 0.4  &
 \\[1.0ex]
\hline
\end{tabular}

\vspace*{0.5cm}

\begin{tabular}{cccc}
\hline\\[-2.0ex]
  & 
 \multicolumn{3}{c}{$[g_1^+]^\rmi{bare}$}
 \\[1.5ex]  
$|\nu|$ & 
 A$_1$ &
 A$_2$ & 
 \cite{prl}   \\[1.5ex]
\hline\\[-2.0ex]
$2$ &
 0.56(13) -- 0.60(14) &
 0.30(9) -- 0.32(10) &
 \\[1.5ex]
$3$  &
 0.47(5) -- 0.51(5)&
 0.53(7) -- 0.57(7)&
 \\[1.0ex]
$4$  &
 0.52(4) -- 0.55(4)&
 0.59(7) -- 0.63(7) &
 \\[1.5ex]
$5$  & 
 0.59(3) -- 0.63(3) &
 0.55(4) -- 0.59(4)&
 \\[1.0ex]
\hline\\[-2.0ex]
w.a. ($|\nu| > 2$) & 
 0.55(2) -- 0.58(2)  &
 0.57(4) -- 0.61(4) &
 0.63(4)\\
 $\chi^2 /{\rm d.o.f.}$  &
  2.4 -- 2.6  &
  0.3 -- 0.2 &
 \\[1.0ex]
\hline
\end{tabular}

\end{center}
\caption{Values of the bare couplings obtained from a NLO fit in the
 interval $|z_0|\leq a$. The ranges indicated correspond 
 to considering $(FL)^2= 1.14-1.40$.}
\label{tab:nlo}
\end{table}
%

The results obtained for $[g_1^\pm]^\rmi{bare}$ on the lattices A$_1$
and A$_2$ are again fully compatible, indicating small scaling violations.
However, the 
effect of a $10\%$  uncertainty in $F$ results in a significant systematic 
uncertainty especially in $[g_1^-]^\rmi{bare}$, which is much more 
sensitive to NLO corrections. In fact the difference between 
the results for $[g_1^-]^\rmi{bare}$ obtained from LO and NLO matchings 
is too large for even the latter results to be trustworthy.

A comparison with the results of a NLO matching in ref.~\cite{prl} shows that 
$[g_1^+]^\rmi{bare}$ is in rather good agreement. This is quite non-trivial, 
given the very different NLO chiral corrections in the two cases, and 
could be an indication that NNLO corrections are not very important 
for this quantity. On the other hand, our $[g_1^-]^\rmi{bare}$ lies 
significantly above the result of ref.~\cite{prl}, but this could be 
accounted for by an effect of $30\%$ in the NNLO corrections, which 
does not appear unreasonable, given that the size of the NLO corrections
is 50 -- 60\%. It is interesting to note, however, that due to the fact that the there is a cancellation between the LO and NLO terms in $R_\nu^-$, the uncertainty in the NLO corrections has a bigger relative  impact in the determination of $[g_1^-]^\rmi{bare}$  than in $[g_1^+]^\rmi{bare}$. 

Obviously it is necessary to 
bring this systematic error under control, which can only be achieved 
by going to larger volumes and performing a detailed finite-size scaling study.
Given that this is a quenched exploratory study and that going to larger 
volumes in the quenched approximation is no guarantee of success, 
we will not pursue this further here. We will consider, however, alternative estimates of
$[g_1^-]^\rmi{bare}$ that could be less affected by higher order corrections.  

One possible strategy\footnote{This was the strategy followed in ref.~\cite{prl}.}  is to obtain $[g_1^-]^\rmi{bare}$ indirectly from $[g_1^+ g_1^-]^\rmi{bare}$
and $[g_1^+]^\rmi{bare}$. The first quantity is extracted from the product $R^+_\nu R^-_\nu$ where  the NLO correction vanishes, while the second quantity is extracted from $R_\nu^+$, where the NLO contribution has the same sign as the LO, and therefore the uncertainty due to higher order effects is less relevant on relative terms.   The results of such an  approach are summarized in Table~\ref{tab:nlo-comb}. The results for $[g_1^-]^\rmi{bare}$ are now very close to those in ref.~\cite{prl}, and are significantly less sensitive to the uncertainty in $F$, as expected. 

\begin{table}[!t]
\small
\begin{center}
\begin{tabular}{cccc}
\hline\\[-2.0ex]
 & 
 \multicolumn{3}{c}{$[g_1^-]^\rmi{bare}$}
 \\[1.5ex]  
$|\nu|$ & 
 A$_1$ &
 A$_2$ &
 \cite{prl}
  \\[1.5ex]
\hline\\[-2.0ex]
$2$ &
 1.3(1.1) -- 1.2(1.0)  &
  2.4(1.1) -- 2.2(1.0)  &
 \\[1.0ex]
$3$  &
 1.99(32) -- 1.85(30) &
 2.1(5) -- 1.9(5)  &
 \\[1.0ex]
$4$  &
 2.66(24) -- 2.47(22) &
 2.71(45)-- 2.54(42) &
 \\[1.0ex]
$5$  & 
 2.79(19) -- 2.60(18) & 
 2.65(24) -- 2.48(22) &
 \\[1.0ex]
\hline\\[-2.0ex]
w.a. ($|\nu| > 2$) & 
 2.61(14) -- 2.42(13)  &
 2.58(20)  -- 2.42(18)&
 2.33(11)
 \\
 $\chi^2 /{\rm d.o.f.}$  &
  2.3 -- 2.3  &
  0.5 -- 0.6 &
 \\[1.0ex]
\hline
\end{tabular}
\end{center}
\caption{Value of the bare coupling $[g_1^-]^\rmi{bare}$ obtained from a NLO fit of $R_\nu^+ R_\nu^-$ and $R_\nu^+$ in the
 interval $|z_0|\leq a$ for lattices A$_1$ and A$_2$. The ranges indicated correspond 
 to considering $(FL)^2= 1.14-1.40$.}
\label{tab:nlo-comb}
\end{table}

Alternatively, as is clear from Table~\ref{tab:rpmq}, on could decrease the chiral corrections very significantly by increasing the distances $\tau_x$ and $\tau_y$ between the pseudoscalar densities and the weak operator, at the expense of decreasing $|\tau_x-\tau_y|$. Note that chiral corrections (up to NLO) are in fact insensitive to the last separation. In principle, the distance between the pseudoscalar sources has to be large enough compared with 
the cutoff and the physical distance scales of QCD. However it is an empirical observation that two-point functions approach the asymptotic behaviour very fast in the $\epsilon$-regime, and we are therefore confident that it makes sense to investigate the three-point functions for larger values of $\tau_x$ and $\tau_y$, even on a lattice as small as $T=16 a$. Provided the effects of higher scales can be neglected, choosing $x_0=7 a$ and $y_0=9 a$ on the lattice A$_1$ can reduce the chiral corrections by 30$\%$. The results obtained for this choice are summarized in Table~\ref{tab:nlo-close}.  

A few observations are in order. The changes in $[g_1^+]^\rmi{bare}$ and  $[g_1^+ g_1^-]^\rmi{bare}$ with respect to the case $\tau_x \simeq \tau_y \simeq |\tau_x-\tau_y|$ are quite small, but the change in $[g_1^-]^\rmi{bare}$ is very significant (yet still at the level expected from NNLO chiral corrections), bringing the value of $[g_1^-]^\rmi{bare}$ to agreement with that from the indirect determination, and with that of ref.~\cite{prl}. The $\chi^2/{\rm d.o.f.}$ of the fits get improved and the effect of the uncertainty in $F$ on  $[g_1^-]^\rmi{bare}$ gets reduced to the level of $10\%$.  In general the agreement of these results with those of  ref.~\cite{prl} is quite remarkable. 
Unfortunately the distance between the sources is  too small to be confident that the effect of higher scales is negligible,  but  these results provide further evidence that the discrepancy between the different determinations of $[g_1^-]^\rmi{bare}$ can indeed be ascribed to higher order chiral corrections.

\begin{table}
\small
\begin{center}
\begin{tabular}{cccc}
\hline\\[-2.0ex]
$|\nu|$ & $[g_1^-]^\rmi{bare}$ & $[g_1^+]^\rmi{bare}$ & $[g_1^+ g_1^-]^\rmi{bare}$ \\
\hline\\[-2.0ex]
$2$ &
 5.5(2.7) -- 4.9(2.4) &
 0.56(14) -- 0.59(14) &
  2.5(1.3)\\[1ex]
$3$  &
 3.45(54) -- 3.10(48)& 
 0.59(7) -- 0.62(7)&
 1.73(35)\\[1ex]
$4$  &
 2.75(31) -- 2.46(28) &
 0.62(5) -- 0.65(5)&
 1.47(23)\\[1ex]
$5$  & 
 2.62(13) -- 2.36(12)&
 0.66(4) -- 0.69(4) &
 1.47(12)\\[1ex]
\hline\\[-2.0ex]
w.a. ($|\nu| > 2$) &
 2.68(11) -- 2.41(11) &
 0.63(3) -- 0.66(3) &
 1.49(10)\\ 
$\chi^2/{\rm d.o.f.}$ &
 1.1 &
 0.45 &
 0.25 
 \\[1.0ex] \hline
\end{tabular}
\end{center}
\caption{Values of the bare couplings obtained from a NLO fit in the
 interval $|z_0|\leq a$ for lattice A$_1$ and $x_0=7 a$, $y_0=9 a$. The ranges indicated correspond 
 to considering $(FL)^2= 1.14-1.40$.}
\label{tab:nlo-close}
\end{table}

%
\section{Conclusions}
\label{sec:conclu}

The purpose of this paper has been to estimate the weak low-energy couplings
$g_1^\pm$, defined in \eq\nr{eq_Hw_xpt}, in the SU(4) chiral limit.
Our method has been to measure the topological zero-mode contributions
to three-point correlation functions of two pseudoscalar densities and 
a weak operator in sectors of non-trivial topology. The results of the 
measurements have been matched to NLO predictions of $\epsilon$-regime
chiral perturbation theory. 

We have considered several fitting strategies for estimating the couplings $g_1^+$ and $g_1^-$, in an attempt to quantify the uncertainty induced by unknown higher order chiral corrections (NNLO), which are expected to be significant at the volume we have considered.  While we observe small variations in the determination of $g_1^+$, as well as in the product $g_1^+ g_1^-$, between the different methods, the value of $g_1^-$ seems to be significantly affected  by higher orders.

Taking the bare couplings $[g_1^+]^\rmi{bare}$ and $[g_1^+ g_1^-]^\rmi{bare}$ cited in Table~\ref{tab:nlo} for the lattice A$_1$, and the same renormalization factors and Wilson coefficients that were
used in ref.~\cite{prl}, and inserting 
everything into \eq\nr{gpm_bare}, we obtain
\be
 g_1^+ \simeq 0.46(5)
 \;, \quad
 g_1^+ g_1^- \simeq 1.2(2)
 \;.
 \label{gpprod}
\ee
The errors shown involve statistical uncertainties as 
well as the uncertainty from the determination of the (quenched) pion decay 
constant $F$. We have checked that discretization
effects in these numbers are small. However, systematic errors related to
higher order chiral corrections as well as the quenched approximation
have not been quantified. In any case, both quantities are in good agreement  
with those of ref.~\cite{prl}, where they were extracted from observables with very different chiral corrections, so this is a strong indication that higher order chiral corrections could be under control.

The situation with $g_1^-$ is less clear. The different fitting strategies we have explored give values that differ by up to 30$\%$, which is also the naive expectation for the magnitude of higher order chiral corrections. We have argued that an indirect extraction of $g_1^-$ from the combinations in eq.~(\ref{gpprod}) is the method that should be least sensitive to the uncertainty induced by higher orders. With this approach, we obtain from the results of Table~\ref{tab:nlo-comb}
 \be
 g_1^- \simeq 2.8(4)
 \;, \quad
  \label{gm}
\ee
which is also in good agreement with the result of ref.~\cite{prl}. We should stress however that we have seen evidence that higher order corrections could be significant for this quantity; the corresponding systematic error cannot be quantified precisely, and has not been included in eq.~(\ref{gm}). Probably an error of $30\%$ in eq.~(\ref{gm}) would be a reasonable estimate.

The conclusions concerning the $\Delta I=1/2$ rule are the same as in ref.~\cite{prl}: there is a significant $\Delta I=1/2$ enhancement already in the SU(4) limit, which cannot be explained by penguin dominance. However, this enhancement is not as large as the experimental one. 
 
The method of the present study and that in refs.~\cite{strategy,prl} can be 
compared on two accounts. First of all, there is the issue of how
well chiral perturbation theory converges with a given box size and
geometry. In a symmetric box of size $(FL)^2 = 1.1$, for instance, 
the magnitude of next-to-leading order corrections in the method of refs.~\cite{strategy,prl}
is $\sim 15\%$ (cf.\ \fig2 of ref.~\cite{strategy}), while in the 
present observables it is $\sim 60\%$ (cf.\ Table \ref{tab:rpmq}).
In the present case the magnitude of the corrections can be reduced very significantly 
by taking the sources further away from the weak operator, 
while in the method of refs.~\cite{strategy,prl} this has no effect.
Nevertheless, it could be concluded that from the point of view of chiral 
perturbation theory, the method of refs.~\cite{strategy,prl} 
appears to be preferable.

The second comparison concerns the numerical cost of the measurements
carried out. On this account, 
on the contrary, the present method appears to be 
preferable: a good statistical signal could be achieved with 
significantly less computational effort than in ref.~\cite{prl}. Indeed, the number of quark propagators required per quark mass to construct the observables is a factor of 5 smaller in the present work. Note also that this factor scales with $N_\rmi{low}$, which is expected to scale with the volume\footnote{A new method  to solve the $V^2$-problem of low-mode preconditioning has been presented in ref.~\cite{deflation}, and in principle could also be applied to low-mode averaging. }.

These two competing aspects probably mean that, moving towards SU(3)
symmetry and unquenched simulations, it would be wise to continue to 
probe the weak low-energy constants 
with (at least) two independent methods. In particular, the fact that chiral corrections are very different in the two cases, offers a good way of quantifying the systematics associated with the chiral fits. On the other hand, particularly in the SU(3) case where the penguin contractions
need to be evaluated, which entails a significant numerical cost, 
it may be that the method introduced in the present work becomes preferable. 

%
\section*{Acknowledgments}
We wish to thank L.~Giusti, M.~L\"uscher and P. Weisz 
for the joint development of important parts of the computer code used in this work.
We acknowledge the computer resources provided by IBM MareNostrum 
at the BSC, the IBM Regatta at FZ J\"ulich and the PC-clusters at 
University of Valencia. P.H.\ and E.T.\ acknowledge partial financial 
support from the research grants FPA-2004-00996, FPA-2007-01678, 
FLAVIAnet and HA2005-0120. C.P. acknowledges financial support
from the Ram\'on y Cajal programme, as well as the research grant FPA2006-05807. C.P. and P. H. acknowledge support from the Consolider-Ingenio 2010 Programme CPAN (CSD2007-00042). 


\appendix
\renewcommand{\thesection}{Appendix~\Alph{section}}
\renewcommand{\thesubsection}{\Alph{section}.\arabic{subsection}}
\renewcommand{\theequation}{\Alph{section}.\arabic{equation}}



%
\section{Graph-by-graph results for the two-point correlator}
\la{app:2pt}

For completeness, we present in this appendix graph-by-graph
results for the two-point correlation function defined in \eq\nr{2pt_xpt}. 

As usual, the Goldstone field is factorised into non-zero 
and zero-mode parts: 
\be
 U(x) \equiv U_\xi(x)\, U_0
 \;, \quad
 U_\xi(x) \equiv \exp\biggl[ \frac{2 i \xi(x)}{F} \biggr]
 \;. 
\ee
The propagator of the non-zero modes $\xi$, 
which are perturbative, is of the form
\be
 \langle \xi_{ij}(x) \, \xi_{kl}(y) \rangle = 
 \fr12 \Bigl[
   \delta_{il}\delta_{jk} G(x-y) - 
   \delta_{ij}\delta_{kl} E(x-y) 
 \Bigr]
 \;. \la{xi_prop} 
\ee
Here $G(x)$ is the massless non-zero mode propagator, 
\be
 G(x) \equiv \frac{1}{V} 
 \sum_{n \in \zz }
 \Bigl(1 - \delta^{(4)}_{n,0} \Bigr) \frac{e^{i p\cdot x}}{p^2} 
 \;, 
 \quad V \equiv T L_1 L_2 L_3
 \;, \quad
 p_0 \equiv \frac{2 \pi n_0}{T}
 \;, \quad 
 p_i \equiv \frac{2\pi n_i}{L_i}
 \;, 
 \la{Gx}
\ee
while $E(x)$ is the ``trace part'' whose form is affected by quenching; in 
the unquenched case it reads $E(x) = G(x)/\Nf$, while in the quenched 
case, 
\be
 E(x) 
  \equiv  \frac{\alpha}{2 N_c} G(x) + \frac{m_0^2}{2 N_c} F(x)
 \;,
 \la{qprop}
\ee
where
\be
 F(x) = \frac{1}{V} 
 \sum_{n \in \zz} 
 \Bigl(1 - \delta^{(4)}_{n,0} \Bigr) \frac{e^{i p \cdot x}}{p^4} 
 \;. \la{Fx}
\ee

Since the zero-mode field $U_0$ is an $x$-independent constant, 
and we are only interested in contributions to the correlation
function $\langle \mathcal{P}^a(x) \mathcal{L}_0^b(z) \rangle_\nu$ 
that remain non-zero after taking
the derivative $\partial_{x_0}$, it is clear that the only graphs
that can contribute are those where the two operators are connected
by a non-zero mode propagator. Denoting the operator $\mathcal{P}^a$
by an open square; the operator $\mathcal{L}^b_0$ by an open half circle; 
the propagator in \eq\nr{xi_prop} by a solid line; 
the mass term in the chiral Lagrangian
by a closed circle; the mass term originating from the Haar measure
by a cross; $\mu\equiv m\Sigma V$; 
and choosing to list the results before taking the spatial average,  
time derivative, and zero-mass limit in \eq\nr{2pt_xpt}, we are led to: 
\ba
 \HTopotree(\TLsc) & = & 
   \frac{i \Sigma}{2} \partial_0 G(z-x) 
  \Bigl\langle
   \tr\Bigl[ (U_0 T^a + T^a U_0^\dagger) T^b \Bigr]
  \Bigr\rangle
 \;,  \la{HTopo_first} \\ 
 \biggl[ \HTopotreeconn(\TLsc,\TAsc)\hspace*{-6mm} \biggr]_\rmi{conn.} & = & 
 - \frac{i \mu\Sigma}{4 F^2} [\Nf G(0) - E(0)]
 \partial_0 G(z-x) 
 \times \nn & & \times
 \biggl[ 
 \Bigl\langle
  \tr[(U_0 T^a + T^a U_0^\dagger) T^b] \tr[U_0 + U_0^\dagger]
 \Bigr\rangle 
 \nn & & - 
 \Bigl\langle
  \tr[(U_0 T^a + T^a U_0^\dagger) T^b]
 \Bigr\rangle 
 \Bigl\langle
    \tr[U_0 + U_0^\dagger]
 \Bigr\rangle 
 \biggr]
 \;,  \\ 
 \HTopomeas(\TLsc,\TLsc) & = &
 - \frac{i \Nf \Sigma}{3 F^2 V}
 \int_s \partial_0 G(z-s) G(s-x) 
 \Bigl\langle
  \tr[(U_0 T^a + T^a U_0^\dagger) T^b]
 \Bigr\rangle   
 \;,  \\
 \HTopomass(\TLsc,\TLsc) & = &
 - \frac{i m\Sigma^2}{4 F^2}  \int_s \partial_0 G(z-s) G(s-x)
 \times \nn & & \times
 \Bigl\langle 
   \tr\Bigl[ 
     (U_0 T^a + T^a U_0^\dagger) \{U_0+U_0^\dagger, T^b \} 
   \Bigr]
 \Bigr\rangle
 \nn & &  
 + \frac{i m\Sigma^2}{2 F^2}  \int_s \partial_0 G(z-s) E(s-x)
 \times \nn & & \times
 \Bigl\langle 
   \tr\Bigl[ 
     U_0 T^a + T^a U_0^\dagger
   \Bigr]
   \tr\Bigl[ 
     T^b (U_0+U_0^\dagger)
    \Bigr]
 \Bigr\rangle
 \;,  \la{eq:HTopomass} \\
 \HTopoin(\TLsc,\TLsc,\TAsc) & = &
 - \frac{i \Nf\Sigma}{6 F^2}
 \Bigl[
  \partial_\nu^2 G(0) \int_s \partial_0 G(z-s) G(s-x)
 \la{HTopoin} \\ & & 
 + G(0) \int_s  \partial_0 \partial_\nu G(z-s) \partial_\nu G(s-x)
 \Bigr] 
  \Bigl\langle
   \tr\Bigl[ (U_0 T^a + T^a U_0^\dagger) T^b \Bigr]
  \Bigr\rangle 
 \;, \nn
 \HTopocucu(\TAsc,\TAsc) & = & 0 
 \;,  \\
 \HTopocu(\TAsc,\TLsc)  & = &
 - \frac{i \Nf  \Sigma}{3 F^2}
 G(0)
 \partial_0 G(z-x) 
  \Bigl\langle
   \tr\Bigl[ (U_0 T^a + T^a U_0^\dagger) T^b \Bigr]
  \Bigr\rangle 
 \;, \\
 \HTopoop(\TAsc,\TLsc)  & = &
 - \frac{i \Sigma}{3 F^2}
 \Bigl[\Nf G(0) - \fr32 E(0) \Bigr]
 \partial_0 G(z-x) 
 \times \nn & & \times
  \Bigl\langle
   \tr\Bigl[ (U_0 T^a + T^a U_0^\dagger) T^b \Bigr]
  \Bigr\rangle 
 \;. \la{HTopo_last}
\ea

In the quenched case, there are two additional contributions, 
discussed around \eqs\nr{Pa_q}, \nr{Znu}:
\ba
 \HTopotreeK(\TLsc) & = & 
   \frac{2 i \nu K \Nc}{m_0^2 F V} \partial_0 G(z-x) \times
  \Bigl\langle
   \tr\Bigl[ (U_0 T^a - T^a U_0^\dagger) T^b \Bigr]
  \Bigr\rangle
 \;,  \hspace*{5mm} \la{HTopo_q_first} \\ 
 \biggl[ \HTopotree(\TLsc) \hspace*{-6mm} \biggr]_\rmi{$K$-weight,conn.} & = & 
   \frac{i \nu m \Sigma K \Nc}{m_0^2 F} \partial_0 G(z-x)
  \times \nn & & \times
  \biggl[ 
  \Bigl\langle
   \tr\Bigl[ (U_0 T^a + T^a U_0^\dagger) T^b \Bigr] \tr(U_0 - U_0^\dagger)
  \Bigr\rangle
  \nn  & & - 
  \Bigl\langle
   \tr\Bigl[ (U_0 T^a + T^a U_0^\dagger) T^b \Bigr]
  \Bigr\rangle
  \Bigl\langle
     \tr(U_0 - U_0^\dagger)
  \Bigr\rangle
  \biggr]
 \;.  \la{HTopo_q_last}  
\ea
Here we made use of the fact that the integral over the 
zero-mode of the singlet field, $\Phi_0$, is Gaussian, and 
$\Phi_0$ can be approximated by the corresponding saddle point value, 
\be
 \Phi_0 = -2 i \frac{\nu\Nc}{m_0^2 F V}
 \;.
\ee
In fact this value was already inserted in order to arrive at 
the $K$-term of \eq\nr{Znu}.

As far as the zero-mode integrals appearing in 
\eqs\nr{HTopo_first}--\nr{HTopo_last} are concerned, the general trick
to use is that, because of the invariance of the integration measure, 
\be
 \langle A_{ij} \, B_{kl} \rangle = 
  c_1\, \delta_{ij} \delta_{kl} + c_2 \, \delta_{il}\delta_{jk}
 \;, \la{trick_1} 
\ee
where $A,B \in \{U_0,U_0^\dagger\}$. 
Carrying out contractions and solving the quadratic system yields
\ba
 c_1 & = & \frac{1}{\Nf(\Nf^2-1)}
 \Bigl\{
 \Nf \langle \tr[A]\; \tr[B] \rangle - 
 \langle \tr[A\, B] \rangle 
 \Bigr\}
 \;, \\ 
 c_2 & = & \frac{1}{\Nf(\Nf^2-1)}
 \Bigl\{
 \Nf \langle \tr[A\, B] \rangle - 
 \langle \tr[A]\; \tr[B] \rangle
 \Bigr\}
 \;. \la{trick_3}
\ea
This leads to integrals whose values
are listed in appendix B of ref.~\cite{zeromode}. For the zero-mass limit 
in \eq\nr{2pt_xpt} we only need the poles in $1/\mu^n$, which are
also listed in ref.~\cite{zeromode}.

Applying this recipe in practice, we obtain 
\ba 
  \lim_{m\to 0} (m V \Sigma)
  \Bigl\langle
   \tr\Bigl[ (U_0 T^a + T^a U_0^\dagger) T^b \Bigr]
  \Bigr\rangle
 & = &  2 |\nu| \tr[T^a T^b]
 \;, \\
 \lim_{m\to 0} (m V \Sigma)^2
 \biggl[ 
 \Bigl\langle
  \tr[(U_0 T^a + T^a U_0^\dagger) T^b] \tr[U_0 + U_0^\dagger]
 \Bigr\rangle  & & 
 \nn - 
 \Bigl\langle
  \tr[(U_0 T^a + T^a U_0^\dagger) T^b]
 \Bigr\rangle 
 \Bigl\langle
    \tr[U_0 + U_0^\dagger]
 \Bigr\rangle 
 \biggr]
  &= &  -4 |\nu| \tr[T^a T^b]
 \;, \\
  \lim_{m\to 0} (m V \Sigma)^2
 \Bigl\langle 
   \tr\Bigl[ 
     (U_0 T^a + T^a U_0^\dagger) \{U_0+U_0^\dagger, T^b \} 
   \Bigr]
 \Bigr\rangle
 & = & 4 |\nu| (2 |\nu| - \Nf ) \tr[T^a T^b]
 \;, \hspace*{5mm} \\ 
   \lim_{m\to 0} (m V \Sigma)^2
 \Bigl\langle 
   \tr\Bigl[ 
     U_0 T^a + T^a U_0^\dagger
   \Bigr]
   \tr\Bigl[ 
     T^b (U_0+U_0^\dagger)
    \Bigr]
 \Bigr\rangle
 & = & - 4 |\nu| \tr[T^a T^b]
 \;. 
\ea
The additional integrals needed in the quenched case
(\eqs\nr{HTopo_q_first}, \nr{HTopo_q_last}) 
read
\ba 
  \lim_{m\to 0} (m V \Sigma)
  \Bigl\langle
   \tr\Bigl[ (U_0 T^a - T^a U_0^\dagger) T^b \Bigr]
  \Bigr\rangle
 & = &  - 2 \nu\, \tr[T^a T^b]
 \;, \la{HTopo_q_zero_first} \\
 \lim_{m\to 0} (m V \Sigma)^2
 \biggl[ 
 \Bigl\langle
  \tr[(U_0 T^a + T^a U_0^\dagger) T^b] \tr[U_0 - U_0^\dagger]
 \Bigr\rangle  & & 
 \nn - 
 \Bigl\langle
  \tr[(U_0 T^a + T^a U_0^\dagger) T^b]
 \Bigr\rangle 
 \Bigl\langle
    \tr[U_0 - U_0^\dagger]
 \Bigr\rangle 
 \biggr]
  &= &  4 \nu\, \tr[T^a T^b]
 \;.\la{HTopo_q_zero_last}
\ea
Inserting the last two into \eqs\nr{HTopo_q_first}, \nr{HTopo_q_last}, 
we see immediately that the two quenched terms cancel against each other. 

Let us finally consider the spacetime dependence. After taking 
the spatial average and time derivative
in \eq\nr{2pt_xpt}, omitting contact terms, and denoting 
$\tau_x = (x_0-z_0)/T$,  we get 
\ba
 \partial_{x_0} \int_\vec{x} 
 \partial_0 G(z-x) & = &  - \frac{1}{T}
 \;, \\
 \partial_{x_0} \int_\vec{x} 
 \int_s \partial_0 G(z-s) G(s-x)
 & = &  T h_1 \bigl(\tau_x\bigr)
 \;, \la{st_A_2} \\ 
 \partial_{x_0} \int_\vec{x} 
 \int_s \partial_0 \partial_\nu G(z-s)  \partial_\nu G(s-x)
 & = & \frac{1}{T}
 \;, 
\ea
where the function $h_1(\tau)$ is given in \eq\nr{h1}.  
In dimensional regularization, the object 
$\partial_\nu^2 G(0)$ appearing in \eq\nr{HTopoin} evaluates to 
$\partial_\nu^2 G(0) =  {1}/{V}$. On the other hand, the 
constants $G(0), E(0)$, appearing in several contributions,
cancel completely in the final result.

Summing now all the results together, but making no assumptions
about the form of $E(x)$, and expressing the result as in \eq\nr{2pt_xpt}, 
we obtain 
\ba
 & & \mathcal{B}_\nu(x_0-z_0)  =  
 \frac{|\nu|}{T}
 \biggl[ 1 + \frac{2|\nu| T^2}{F^2 V} h_1(\tau_x) + 
 \frac{2 T}{F^2 V}  
 \partial_{x_0} \int_\vec{x} \int_s \partial_0 G(z-s) E(s-x)
 \biggr]
 \;. \hspace*{1cm} \la{B_complete}
\ea

%
\section{Graph-by-graph results for the three-point correlator}
\la{app:3pt}

For completeness, we present in this appendix graph-by-graph
results for the three-point correlation function defined in \eq\nr{3pt_xpt}. 
To be precise, we list results for the operator $\mathcal{O}_{rsuv}$ from 
\eq\nr{Orsuv_def_xpt}, with $r,s,u,v$ assumed to be all different; results
for the operator $\mathcal{O}_1$ are then obtained by symmetrizing
according to \eq\nr{O1_def_xpt}. 

\def\TopoLaction(#1,#2,#3){\piccc{#1(0,15)(16,15) #2(24,15)(40,15)%
#3(40,15)(80,15) %
\SetWidth{1.0} \Line(16,15)(20,19) \Line(20,19)(24,15)%
     \Line(24,15)(20,11) \Line(20,11)(16,15) \SetWidth{1.0}%
\GBoxc(0,15)(5,5){1} \GBoxc(80,15)(5,5){1} \GCirc(40,15){3}{1} }}
\def\TopoLcurrent(#1,#2){\piccc{#1(0,15)(40,15)%
#2(40,15)(80,15) %
\GBoxc(0,15)(7.5,7.5){1} \GBoxc(80,15)(5,5){1} \GCirc(40,15){3}{1}%
\SetWidth{1.0} \Line(-3.5,15)(0,18.5) \Line(0,18.5)(3.5,15)%
     \Line(3.5,15)(0,11.5) \Line(0,11.5)(-3.5,15) \SetWidth{1.0}%
}}
\def\TopoLoperator(#1,#2){\piccc{#1(0,15)(40,15)%
#2(40,15)(80,15) %
\GBoxc(0,15)(5,5){1} \GBoxc(80,15)(5,5){1} \GCirc(40,15){5}{1}%
\SetWidth{1.0} \Line(36,15)(40,19) \Line(40,19)(44,15)%
     \Line(44,15)(40,11) \Line(40,11)(36,15) \SetWidth{1.0}%
 }}

Since the zero-mode field $U_0$ is an $x$-independent constant, 
and we are only interested in contributions to the correlation
function 
$
 - \langle
  \mathcal{P}^a(x) 
  \mathcal{O}_{rsuv}(z) 
  \mathcal{P}^b(y)
  \rangle_\nu
$ 
that remain non-zero after taking
the derivatives $\partial_{x_0}\partial_{y_0}$, it is clear that the only 
graphs that can contribute are those where the pseudoscalar densities are 
connected to each other or to the weak operator by non-zero mode propagators. 
Furthermore, in the SU(4) limit we can assume
all the indices $\a,\b,\c,\d$ to be different, which allows us to omit 
structures like $\delta_{ur}, \delta_{us}, \delta_{vr}, \delta_{vs}$; 
this means that the weak operator needs to be connected to at least 
one of the pseudoscalar densities. 
Denoting the operator $\mathcal{P}^a$ by an open square; 
the operator $\mathcal{O}_{\a\b\c\d}$ by an open circle; 
the propagator in \eq\nr{xi_prop} by a solid line; and choosing to list 
the results before taking the spatial averages, time derivatives, 
and zero-mass limit in \eq\nr{3pt_xpt}, 
we are led to: 
\ba
 \Topotree(\TLsc,\TLsc) 
 & = &
 \frac{\Sigma^2}{4}
 \partial_\mu G(x-z) \partial_\mu G(y-z) \times 
 \nn & & \times
 \biggl[
  \Bigl\langle
    (U_0 T^a + T^a U_0^\dagger)_{ur} 
    (U_0 T^b + T^b U_0^\dagger)_{vs} 
  \Bigr\rangle 
  + (a\leftrightarrow b)
 \biggr] 
 \;, \hspace*{5mm} \\
 \biggl[\Topotreemass(\TLsc,\TLsc,\TAsc) \!\!\! \biggr]_\rmi{conn.} 
 & = & 
 \frac{\mu\Sigma^2}{8 F^2} [\Nf G(0) - E(0) ]
 \partial_\mu G(x-z) \partial_\mu G(y-z) \times 
 \nn & & \times
 \biggl[
  \Bigl\langle
    (U_0 T^a + T^a U_0^\dagger)_{ur} 
    (U_0 T^b + T^b U_0^\dagger)_{vs} 
  \Bigr\rangle \Bigl\langle \tr(U_0 + U_0^\dagger) \Bigr\rangle 
 \nn & & 
 - \Bigl\langle
    (U_0 T^a + T^a U_0^\dagger)_{ur} 
    (U_0 T^b + T^b U_0^\dagger)_{vs} 
     \tr(U_0 + U_0^\dagger) \Bigr\rangle 
 \nn & & 
  + (a\leftrightarrow b)
 \biggr]  
 \;, \\
 \Topomeas(\TLsc,\TLsc,\TLsc) 
 & = & - \frac{\Nf \Sigma^2}{6 F^2 V}
 \biggl[
    \partial_\mu G(x-z) \int_s \partial_\mu G(y-s) G(s-z)
    + (x\leftrightarrow y)
 \biggr] \times
 \nn & & \times
 \biggl[
  \Bigl\langle
    (U_0 T^a + T^a U_0^\dagger)_{ur} 
    (U_0 T^b + T^b U_0^\dagger)_{vs} 
  \Bigr\rangle 
  + (a\leftrightarrow b)
 \biggr]  
 \;, \\
 \Topomass(\TLsc,\TLsc,\TLsc) 
 & = & -\frac{m\Sigma^3}{8 F^2} 
  \times  \biggl\{ 
    \partial_\mu G(x-z) \int_s \partial_\mu G(y-s) G(s-z) \times
 \nn & & \times 
  \Bigl\langle
    (U_0 T^a + T^a U_0^\dagger)_{\c\a} 
    \{ U_0 +U_0^\dagger, 
       U_0 T^b + T^b U_0^\dagger \}_{\d\b} 
  \Bigr\rangle 
 \nn &  & 
  + \partial_\mu G(y-z) \int_s \partial_\mu G(x-s) G(s-z) \times
 \nn & & \times 
  \Bigl\langle
    (U_0 T^b + T^b U_0^\dagger)_{\c\a} 
    \{ U_0 +U_0^\dagger, 
       U_0 T^a + T^a U_0^\dagger \}_{\d\b} 
  \Bigr\rangle 
 \nn &  & 
   - 2 \partial_\mu G(x-z) \int_s \partial_\mu E(y-s) G(s-z) \times
 \nn & & \times 
  \Bigl\langle
    (U_0 T^a + T^a U_0^\dagger)_{\c\a} 
    (U_0 +U_0^\dagger)_{\d\b} 
    \tr(U_0 T^b + T^b U_0) 
  \Bigr\rangle 
 \nn &  & 
  - 2 \partial_\mu G(y-z) \int_s \partial_\mu E(x-s) G(s-z) \times
 \nn & & \times 
  \Bigl\langle
    (U_0 T^b + T^b U_0^\dagger)_{\c\a} 
     (U_0 +U_0^\dagger)_{\d\b} 
     \tr(U_0 T^a + T^a U_0^\dagger) 
  \Bigr\rangle 
 \nn 
 & & + (\a\leftrightarrow\b,\c\leftrightarrow\d) \biggr\} 
 \;, \\
 \Topoin(\TLsc,\TLsc,\TLsc,\TAsc) 
 & = & -\frac{\Nf \Sigma^2}{12 F^2}
 \biggl[
     G(0) \partial_\mu G(x-z) 
     \int_s \partial_\mu \partial_\nu G(y-s)  \partial_\nu G(s-z)
   \nn & & + 
     \partial_\nu^2 G(0) \partial_\mu G(x-z) 
     \int_s \partial_\mu G(y-s)  G(s-z)
   + (x\leftrightarrow y) 
 \biggr] \times
 \nn & & \times 
 \biggl[
  \Bigl\langle
    (U_0 T^a + T^a U_0^\dagger)_{\c\a} 
    (U_0 T^b + T^b U_0^\dagger)_{\d\b} 
  \Bigr\rangle 
  + (a\leftrightarrow b)
  \biggr]
 \;, \\
 \Topoinop(\TLsc,\TLsc,\TAsc) 
 & = & -\frac{\Sigma^2}{12 F^2}
 \int_s \biggl[
     G(x-s) G(y-s) 
     \partial_\mu \partial_\nu G(s-z) 
     \partial_\mu \partial_\nu G(s-z)
   \nn & & + 
     \partial_\nu G(x-s) G(y-s) 
     \partial_\mu  G(s-z) 
     \partial_\mu \partial_\nu G(s-z)
   \nn & & + 
     G(x-s) \partial_\nu G(y-s) 
     \partial_\mu  G(s-z) 
     \partial_\mu \partial_\nu G(s-z)
   \nn & & + 
     \partial_\nu G(x-s) \partial_\nu G(y-s) 
     \partial_\mu  G(s-z) 
     \partial_\mu  G(s-z)
 \biggr] \times
 \nn & & \times 
 \biggl[
  \Bigl\langle
    (U_0 T^a + T^a U_0^\dagger)_{\c\b} 
    (U_0 T^b + T^b U_0^\dagger)_{\d\a} 
  \Bigr\rangle 
  + (a\leftrightarrow b)
  \biggr]
 \;, \\
 \Topocu(\TAsc,\TLsc,\TLsc) 
 & = & 
 \frac{\Sigma^2}{3 F^2} \biggl[\fr32 E(0) - \Nf G(0)\biggr]
 \partial_\mu G(x-z) \partial_\mu G(y-z) \times 
 \nn & & \times
 \biggl[
  \Bigl\langle
    (U_0 T^a + T^a U_0^\dagger)_{ur} 
    (U_0 T^b + T^b U_0^\dagger)_{vs} 
  \Bigr\rangle 
 + (a\leftrightarrow b)
 \biggl] 
 \;, \\
 \Topocucuop(\TAsc,\TAsc,\TAsc) 
 & = & 
 - \frac{\Sigma^2}{12 F^2}
 \biggl\{
   G(x-y) \Bigl[ \partial_\mu G(y-z)\Bigr]^2 
  + (x\leftrightarrow y) 
 \biggr\} \times
 \nn & & \times
 \biggl[
  \Bigl\langle
    (U_0 T^a + T^a U_0^\dagger)_{\c\b} 
    (U_0 T^b + T^b U_0^\dagger)_{\d\a} 
  \Bigr\rangle 
 + (a\leftrightarrow b)
 \biggl] 
  \;, \\
 \Topocucu(\TAsc,\TLsc,\TLsc) 
 & = & -\frac{\Sigma^2}{4 F^2}
   G(x-y) \partial_\mu G(x-z) \partial_\mu G(y-z) 
 \times \nn & & \times
 \biggl[
  \Bigl\langle
    (U_0 T^a - T^a U_0^\dagger)_{\c\b} 
    (U_0 T^b - T^b U_0^\dagger)_{\d\a} 
  \Bigr\rangle 
 + (a\leftrightarrow b)
 \biggl] 
 \nn & & + \frac{\Sigma^2}{2 F^2}
   E(x-y) \partial_\mu G(x-z) \partial_\mu G(y-z) 
 \times \nn & & \times
 \biggl[
  \Bigl\langle
    (U_0 T^a - T^a U_0^\dagger)_{\c\a} 
    (U_0 T^b - T^b U_0^\dagger)_{\d\b} 
  \Bigr\rangle 
 + (a\leftrightarrow b)
 \biggl]
 \;, \la{eq:Topocucu} \\
 \Topocuop(\TAsc,\TAsc,\TLsc) 
 & = & 0 
 \;, \\
 \Topocuopbub(\TLsc,\TAsc,\TAsc)
 & = & 0 
 \;, \\
 \Topoop(\TAsc,\TLsc,\TLsc) 
 & = & -\frac{\Sigma^2}{12 F^2} 
 \Bigl[ \partial_\mu^2 G(0) G(x-z)G(y-z) 
 \nn & & + 
 3 G(0) \partial_\mu G(x-z) \partial_\mu G(y-z) \Bigr]
 \times \nn & & \times
 \biggl[
  \Bigl\langle
    (U_0 T^a + T^a U_0^\dagger)_{\c\b} 
    (U_0 T^b + T^b U_0^\dagger)_{\d\a} 
  \Bigr\rangle 
 + (a\leftrightarrow b)
 \biggl] 
 \nn & &  -\frac{\Nf \Sigma^2}{3 F^2}
 G(0) \partial_\mu G(x-z) \partial_\mu G(y-z)
 \times \nn & & \times
 \biggl[
  \Bigl\langle
    (U_0 T^a + T^a U_0^\dagger)_{\c\a} 
    (U_0 T^b + T^b U_0^\dagger)_{\d\b} 
  \Bigr\rangle 
 + (a\leftrightarrow b)
 \biggl] 
 \;.
\ea
In the quenched case, there are two additional contributions,
\ba
 \TopotreesmallK(\TLsc,\TLsc) 
 & = &
 \frac{\nu\Sigma K \Nc}{m_0^2 F V}\;
 \partial_\mu G(x-z) \partial_\mu G(y-z) \times 
 \nn & & \times
 \biggl[
  \Bigl\langle
    (U_0 T^a + T^a U_0^\dagger)_{\c\a} 
    (U_0 T^b - T^b U_0^\dagger)_{\d\b} 
  \Bigr\rangle 
 \nn & & 
  + \Bigl\langle
    (U_0 T^a + T^a U_0^\dagger)_{\d\b} 
    (U_0 T^b - T^b U_0^\dagger)_{\c\a} 
  \Bigr\rangle 
 \nn & & 
  + (a\leftrightarrow b)
 \biggr] 
 \;, \hspace*{5mm}  \la{Topo_q_first} \\
 \biggl[ \Topotreesmall(\TLsc,\TLsc)  \biggr]_\rmi{$K$-weight,conn.} 
 & = &
 \frac{\nu m \Sigma^2 K \Nc}{2 m_0^2 F} \;
 \partial_\mu G(x-z) \partial_\mu G(y-z) \times 
 \nn & & \times
 \biggl[
  \Bigl\langle
    (U_0 T^a + T^a U_0^\dagger)_{ur} 
    (U_0 T^b + T^b U_0^\dagger)_{vs} \tr(U_0 -U_0^\dagger) 
  \Bigr\rangle 
 \nn & & - 
  \Bigl\langle
    (U_0 T^a + T^a U_0^\dagger)_{ur} 
    (U_0 T^b + T^b U_0^\dagger)_{vs}
  \Bigr\rangle
  \Bigl\langle
    \tr(U_0 -U_0^\dagger) 
  \Bigr\rangle 
 \nn & & 
  + (a\leftrightarrow b)
 \biggr] 
 \;.  \la{Topo_q_last}
\ea

Employing the same trick as in \eqs\nr{trick_1}--\nr{trick_3}, 
the zero-mode integrals become
\ba
 & &  \lim_{m\to 0} (m V \Sigma)^2
 \biggl\{
  \Bigl\langle
    (U_0 T^a + T^a U_0^\dagger)_{ur} 
    (U_0 T^b + T^b U_0^\dagger)_{vs}  
 \Bigr\rangle + (a \leftrightarrow b)
 \biggr\}
 \nn 
 &  & \hspace*{2cm} =  4 \Bigl[ \nu^2 T^{\{a}_{ur}T^{b\}}_{vs} 
         - |\nu| T^{\{a}_{us}T^{b\}}_{vr} \Bigr] 
 \;, \\  
 &  &  \lim_{m\to 0} (m V \Sigma)^2
 \biggl\{
  \Bigl\langle
    (U_0 T^a - T^a U_0^\dagger)_{ur} 
    (U_0 T^b - T^b U_0^\dagger)_{vs}  
 \Bigr\rangle + (a \leftrightarrow b)
 \biggr\}
 \nn 
 &  & \hspace*{2cm} =  4 \Bigl[ \nu^2 T^{\{a}_{ur}T^{b\}}_{vs} 
         - |\nu| T^{\{a}_{us}T^{b\}}_{vr} \Bigr] 
 \;, \\  
 & &  \lim_{m\to 0} (m V \Sigma)^3
 \biggl\{
  \Bigl\langle
    (U_0 T^a + T^a U_0^\dagger)_{ur} 
    (U_0 T^b + T^b U_0^\dagger)_{vs} 
  \Bigr\rangle \Bigl\langle \tr(U_0 + U_0^\dagger) \Bigr\rangle 
 \nn & & \hspace*{2cm}
 - \Bigl\langle
    (U_0 T^a + T^a U_0^\dagger)_{ur} 
    (U_0 T^b + T^b U_0^\dagger)_{vs} 
     \tr(U_0 + U_0^\dagger) \Bigr\rangle 
  + (a\leftrightarrow b)
 \biggr\}  
 \nn 
 &  & \hspace*{2cm} =   16 \Bigl[ \nu^2 T^{\{a}_{ur}T^{b\}}_{vs} 
         - |\nu| T^{\{a}_{us}T^{b\}}_{vr} \Bigr] 
 \;, \\
 & &  \lim_{m\to 0} (m V \Sigma)^3
 \biggl\{
  \Bigl\langle
    (U_0 T^a + T^a U_0^\dagger)_{\c\a} 
    \{ U_0 +U_0^\dagger, 
       U_0 T^b + T^b U_0^\dagger \}_{\d\b} 
  \Bigr\rangle 
  + (\a\leftrightarrow\b,\c\leftrightarrow\d)
 \biggr\}  
 \nn 
 &  & \hspace*{2cm}  = 8 \Bigl[ (|\nu| - \Nf \nu^2 + 2 \nu^2 |\nu| )
         T^{\{a}_{\c\a}T^{b\}}_{\d\b} 
         + (\Nf |\nu| - 3 \nu^2) 
         T^{\{a}_{\c\b}T^{b\}}_{\d\a} 
           \Bigr] 
 \;, \\
 & &  \lim_{m\to 0} (m V \Sigma)^3
 \biggl\{
   \Bigl\langle
    (U_0 T^a + T^a U_0^\dagger)_{\c\a} 
     (U_0 +U_0^\dagger)_{\d\b} 
     \tr(U_0 T^b + T^b U_0^\dagger) 
  \Bigr\rangle 
  + (\a\leftrightarrow\b,\c\leftrightarrow\d)
 \biggr\}  
 \nn 
 &  & \hspace*{2cm} =   8 \Bigl[ - \nu^2 T^{\{a}_{\c\a}T^{b\}}_{\d\b} 
         + |\nu| T^{\{a}_{\c\b}T^{b\}}_{\d\a} \Bigr] 
 \;,
\ea
where we again omitted all terms proportional to 
$\delta_{us}, \delta_{ur}, \delta_{vs}, \delta_{vr}$.

The additional zero-mode integrals needed in the quenched case
read
\ba
 & &  \lim_{m\to 0} (m V \Sigma)^2
 \biggl\{
  \Bigl\langle
    (U_0 T^a + T^a U_0^\dagger)_{\c\a} 
    (U_0 T^b - T^b U_0^\dagger)_{\d\b} 
  \Bigr\rangle 
 \nn & & \hspace*{2cm}
  + \Bigl\langle
    (U_0 T^a + T^a U_0^\dagger)_{\d\b} 
    (U_0 T^b - T^b U_0^\dagger)_{\c\a} 
  \Bigr\rangle 
 + (a \leftrightarrow b)
 \biggr\}
 \nn 
 &  & \hspace*{2cm} =  -8 \nu \Bigl[  |\nu|  T^{\{a}_{ur}T^{b\}}_{vs} 
         - T^{\{a}_{us}T^{b\}}_{vr} \Bigr] 
 \;, \la{Topo_zm_q_first} \\  
 & &  \lim_{m\to 0} (m V \Sigma)^3
 \biggl\{
  \Bigl\langle
    (U_0 T^a + T^a U_0^\dagger)_{ur} 
    (U_0 T^b + T^b U_0^\dagger)_{vs} 
   \tr(U_0 - U_0^\dagger) \Bigr\rangle 
 \nn & & \hspace*{2cm}
 - \Bigl\langle
    (U_0 T^a + T^a U_0^\dagger)_{ur} 
    (U_0 T^b + T^b U_0^\dagger)_{vs} 
  \Bigr\rangle \Bigl\langle   \tr(U_0 - U_0^\dagger) \Bigr\rangle 
  + (a\leftrightarrow b)
 \biggr\}  
 \nn 
 &  & \hspace*{2cm} =   16 \nu \Bigl[ |\nu| T^{\{a}_{ur}T^{b\}}_{vs} 
         -  T^{\{a}_{us}T^{b\}}_{vr} \Bigr] 
 \;. \la{Topo_zm_q_last}
\ea
Inserting the last two into \eqs\nr{Topo_q_first}, \nr{Topo_q_last}, 
we see immediately that terms proportional to $K$ cancel against each other. 

Let us finally consider the spacetime dependence. After taking the
spatial averages and time derivatives in \eq\nr{3pt_xpt}, omitting 
contact terms, and denoting 
$\tau_x = (x_0-z_0)/T$,  
$\tau_y = (y_0-z_0)/T$,  
we get 
\ba
 & & \partial_{x_0} \int_\vec{x} 
 G(x-z) =  h_1'\bigl(\tau_x\bigr)
 \;, \\
 & & \partial_{x_0} \int_\vec{x} 
 \partial_\mu G(x-z) =  \frac{\delta_{\mu 0}}{T}
 \;, \\
 & &  \partial_{y_0} \int_\vec{y} 
 \int_s \partial_\mu G(y-s) G(s-z)
 =   - \delta_{\mu 0} T h_1 \bigl(\tau_y\bigr)
 \;, \\ 
 & & \partial_{y_0} \int_\vec{y} 
 \int_s \partial_\mu \partial_\nu G(y-s)  \partial_\nu G(s-z)
  =  -\frac{\delta_{\mu 0}}{T}
 \;, \\ 
 & & \partial_{x_0} \partial_{y_0} 
 \int_\vec{x,y}
  G(x-y) \Bigl[ \partial_\mu G(y-z)\Bigr]^2 
  \nn & & \hspace*{2cm} = \frac{1}{V} 
 \Bigl\{ 
   -f_1\bigl( \tau_y \bigr)
  +2 h_1'\bigl( \tau_x - \tau_y \bigr)
  \Bigl[ 
    h_1'\bigl( \tau_y \bigr)
   + 2 \sum_{\vec{p}\neq \vec{0}} |\vec{p}|^2 
   C_\vec{p}\bigl( \tau_y \bigr)
   C_\vec{p}'\bigl( \tau_y \bigr)
  \Bigr] 
 \Bigr\}
 \;, \\ 
 & & \partial_{x_0} \partial_{y_0} 
 \int_\vec{x,y}
   G(x-y) \partial_\mu G(x-z) \partial_\mu G(y-z) 
 \nn & & \hspace*{2cm} = \frac{1}{V} 
 \Bigl\{
   - h_1'(\tau_x) h_1'(\tau_y) + 
     h_1'(\tau_x-\tau_y)
     \Bigl[ h_1'(\tau_x)  - h_1'(\tau_y) \Bigr]
   + h_1 (\tau_x - \tau_y)
 \Bigr\} \hspace*{1cm}
 \\ &  & \hspace*{2cm} = -\frac{1}{V} H(\tau_x,\tau_y) 
 \;,  \la{eq:GGGint} \\
 & & \partial_{x_0} \partial_{y_0} 
 \int_\vec{x,y}
  \int_s \biggl[  
     G(x-s) G(y-s) 
     \partial_\mu \partial_\nu G(s-z) 
     \partial_\mu \partial_\nu G(s-z)
    \nn &  & \hspace*{2cm} +
     \partial_\nu G(x-s) G(y-s) 
     \partial_\mu  G(s-z) 
     \partial_\mu \partial_\nu G(s-z)
    \nn &  & \hspace*{2cm} +
     G(x-s) \partial_\nu G(y-s) 
     \partial_\mu  G(s-z) 
     \partial_\mu \partial_\nu G(s-z)
    \nn &  & \hspace*{2cm} + 
     \partial_\nu G(x-s) \partial_\nu G(y-s) 
     \partial_\mu  G(s-z) 
     \partial_\mu  G(s-z)
 \biggr]
  \nn & & \hspace*{2cm} =  3 \frac{G(0)}{T^2} + 
 \frac{1}{2V} 
 \Bigl\{
     h_1'(\tau_x) h_1'(\tau_y) 
   + 7 h_1'(\tau_x-\tau_y)
     \Bigl[ h_1'(\tau_x)  - h_1'(\tau_y) \Bigr]
   \nn & & \hspace*{2cm}
   - 3 \Bigl[  h_1 (\tau_x - \tau_y) + 
    h_1 (\tau_x) +  h_1 (\tau_y) 
    \Bigr] -4 \Bigl[ f_1(\tau_x) + f_1(\tau_y) \Bigr] 
 \Bigr\} 
   \nn & & \hspace*{2cm}
 -\frac{4}{V} 
 \sum_{\vec{p}\neq\vec{0}} 
 \Bigl\{  
  h_1'(\tau_x-\tau_y) |\vec{p}|^2 
  \Bigl[ 
   C_\vec{p}\bigl( \tau_y \bigr)
   C_\vec{p}'\bigl( \tau_y \bigr) 
  -
   C_\vec{p}\bigl( \tau_x \bigr)
   C_\vec{p}'\bigl( \tau_x \bigr) 
  \Bigr]
 \Bigr\}
 \;.
\ea
The functions $h_1$, $f_1$, $C_\vec{p}$ appearing here have been defined
in \eqs\nr{h1}, \nr{f1}, \nr{Cp}, respectively, and in \eq\nr{eq:GGGint}
we identified the function $H$ defined in \eq\nr{Hxy}. 
We also need to know that in dimensional regularization, 
$\partial_\mu^2 G(0)  =  {1}/{V}$ and 
\be
 G(0) =  -\frac{\beta_1}{\sqrt{V}}
 \;, \la{G0} 
\ee 
where $\beta_1$ is a ``shape coefficient''~\cite{hal,h}.
It is furthermore useful to note the identity 
\be
   h_1'(\tau_x) h_1'(\tau_y) 
   + h_1'(\tau_x-\tau_y)
     \Bigl[ h_1'(\tau_x)  - h_1'(\tau_y) \Bigr] = 
    h_1 (\tau_x - \tau_y) + 
    h_1 (\tau_x) +  h_1 (\tau_y) 
  \;. 
\ee

Summing all the results together, but making no assumptions about
the form of $E(x)$, we obtain
\ba
 & & \hspace*{-1.5cm} \lim_{m\rightarrow 0} 
 (mV)^2 T^2 \partial_{x_0} \partial_{y_0}  
 \int_{\vec{x}} 
 \int_{\vec{y}}\Bigl\langle 
 - \mathcal{P}^a(x) 
 \mathcal{O}_{rsuv}(z) 
 \mathcal{P}^b(y)
 \Bigr\rangle_\nu
 \nn 
 & = &
 \Bigl[ \nu^2 T^{\{a}_{ur}T^{b\}}_{vs} 
      - |\nu| T^{\{a}_{us}T^{b\}}_{vr} \Bigr]
 \times \biggl\{
 1 +  \frac{2|\nu| T^2}{F^2 V} 
 \Bigl[ h_1(\tau_x) + h_1(\tau_y) \Bigr] 
 \nn & & \hspace*{1cm}
 + \frac{2 T}{F^2 V} 
 \Bigl[  
 \partial_{x_0} \int_\vec{x} \int_s \partial_0 G(z-s) E(s-x)
 + 
 \partial_{y_0} \int_\vec{y} \int_s \partial_0 G(z-s) E(s-y)
 \Bigr] 
 \nn & &  \hspace*{1cm}
 + \frac{2 T^2}{F^2} 
 \partial_{x_0} \partial_{y_0} 
 \int_\vec{x,y}
   E(x-y) \partial_\mu G(x-z) \partial_\mu G(y-z)
 \biggr\}
 \nn & + & 
 \Bigl[- |\nu| T^{\{a}_{ur}T^{b\}}_{vs} 
       +  \nu^2 T^{\{a}_{us}T^{b\}}_{vr} \Bigr]
 \times \biggl\{
 -\frac{2 G(0)}{F^2}
 \nn & &  \hspace*{1cm}
 + \frac{T^2}{F^2 V} 
 \Bigl[
  f_1(\tau_x) + f_1 (\tau_y) - h_1(\tau_x) - h_1(\tau_y) + H(\tau_x,\tau_y) 
 \Bigr]
 \biggr\}
 \;. \la{A_complete_1}
\ea
Taking finally the combinations in \eq\nr{O1_def_xpt}; 
writing the result in the form of \eq\nr{Anu_pm_xpt}; 
identifying expressions of the form in \eq\nr{B_complete} from 
the result; and inserting \eq\nr{G0} as well as the definition 
$\rho\equiv T/L$, we obtain
\ba
 & & \hspace*{-1.5cm}
 \bar\mathcal{A}_\nu(x_0-z_0,y_0-z_0) \pm 
 \tilde\mathcal{A}_\nu(x_0-z_0,y_0-z_0)
 \nn  & = & 
 \Bigl( 1 \mp \frac{1}{|\nu|} \Bigr)
 \biggl\{
   \mathcal{B}_\nu(x_0-z_0)  
   \mathcal{B}_\nu(y_0-z_0)
 \nn & & \hspace*{1cm}  
 + \frac{2 \nu^2}{F^2} 
 \partial_{x_0} \partial_{y_0} 
 \int_\vec{x,y}
   E(x-y) \partial_\mu G(x-z) \partial_\mu G(y-z)
 \nn & &  \hspace*{1cm}
 \pm \frac{\nu^2}{F^2 V}
 \Bigl[ 
  2 \beta_1 \rho^{-\fr32} 
 + 
  f_1(\tau_x) + f_1 (\tau_y) - h_1(\tau_x) - h_1(\tau_y) + H(\tau_x,\tau_y) 
 \Bigr]  
 \biggr\}
 \;.  \hspace*{0.5cm} \la{A_complete_2}
\ea


\end{document}